\documentclass[a4paper,11pt]{article}
\usepackage{jheppub}

\usepackage{caption}
\usepackage{lineno}
\usepackage{indentfirst}
 \usepackage{lmodern}
 \usepackage{graphics}
 \usepackage{float}
 \usepackage{tikz-cd} 
\usepackage{mathrsfs}
\usepackage{tikz}
\tikzset{transition/.style={->,line width=1pt}}
\usepackage{url}
\usepackage{float}
\usepackage{standalone}
\usepackage{bm}
\usepackage{amssymb}

\usepackage{cleveref}
\usepackage{subcaption}\usepackage{longtable}
\usepackage{booktabs}
\usepackage{graphicx} 
\usepackage{array}
\usepackage{paralist}
\usepackage{amsmath}
\usepackage{physics}
\usepackage{float}
\usepackage{xcolor}
\usepackage{amsfonts}
\usepackage{amssymb}
\usepackage{verbatim}
 \usepackage{indentfirst}
\usepackage{tikz}
\usepackage{soul}
\usepackage{pifont}
\usepackage{forest}
 
\usepackage{tikz-cd}  
\usetikzlibrary{calc}
 \usepackage{tikz}
\usepackage{multirow} 
\usepackage{makecell}
\usepackage{longtable}

\title{ADE Quiver Duality Proposals: Graph Classification and Nodewise Realizations}
\author{Yuanyuan Fang}
\affiliation[]{Shing-Tung Yau Center and School of Physics, Southeast University, Nanjing 210096, China}
\emailAdd{yuanyuan\_fang@seu.edu.cn}
 
\abstract{We use the large-$N$ index strategy to find new quiver dualities in four-dimensional $\mathcal N=1$ quiver gauge theories. We study gauge quivers with or without an adjoint field at each gauge node, and bifundamental matter between selected nodes with fixed $R$-charges. Imposing the finite-meson condition restricts the allowed gauge graphs to finite ADE diagrams and leads to a new simultaneous duality for ADE quivers. We present an explicit superpotential that produces a uniform classical meson spectrum consistent with the large-$N$ index approach's prediction. We then relate the simultaneous rank transformation to the longest element of the ADE Weyl group, whose reduced expressions organize nodewise duality sequences realizing the simultaneous ADE duality. Realizing the simultaneous ADE duality through such nodewise sequences offers a finite-$N$ consistency check beyond the large-$N$ index construction. Moreover, we realize the simultaneous ADE duality as a sequence of subquiver simultaneous dualities.}

\begin{document} 
\maketitle
\flushbottom

\section{Introduction}
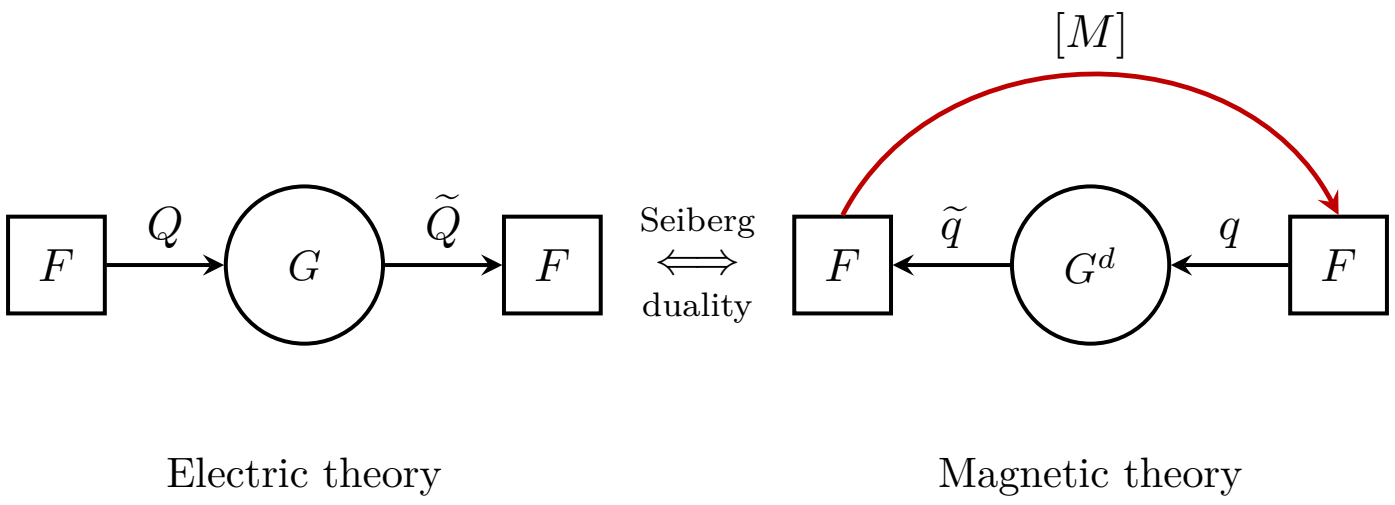
\begin{figure}[tb]
  \centering
  \begin{tikzpicture}[
    scale=0.8, transform shape,
    line width=0.75pt, >=stealth,
    gauge/.style={circle,draw,minimum size=13mm,inner sep=1pt,
      align=center,font=\small},
    flavor/.style={rectangle,draw,minimum size=8mm,inner sep=1pt},
    meson/.style={draw=red!75!black,line width=0.9pt}]
    \node[gauge] (EG) at (-3.25,0)
      {$G$};
    \node[flavor] (EFi) at (-5.30,0) {$F$};
    \node[flavor] (EFj) at (-1.20,0) {$F$};
    \draw[->] (EFi.east) to[bend left=0]
      node[midway,above] {$Q$} (EG.west);
    \draw[->] (EG.east) to[bend left=0]
      node[midway,above] {$\widetilde Q$} (EFj.west);
   \node[font=\small] at (-3.25,-1.75) {Electric theory};

    \node at (0,0) {$\Longleftrightarrow$};
    \node[font=\scriptsize] at (0,0.35) {Seiberg};
\node[font=\scriptsize] at (0,-0.35) {duality};
    \node[gauge] (MG) at (3.25,0)
      {$G^d$};
    \node[flavor] (MFi) at (1.20,0) {$F$};
    \node[flavor] (MFj) at (5.30,0) {$F$};
    \draw[<-] (MFi.east) to[bend left=0]
      node[midway,above] {$\widetilde q$} (MG.west);
    \draw[<-] (MG.east) to[bend left=0]
      node[midway,above] {$q$} (MFj.west);
    \draw[meson,->] (MFi.north)
      .. controls (2.0,1.95) and (4.5,1.95) ..
      node[midway,above] {$[M]$} (MFj.north);

    \node[font=\small] at (3.25,-1.75) {Magnetic theory};
  \end{tikzpicture}
  \caption{Original Seiberg duality for SQCD. The electric theory has gauge group $G=SU(N)$,
$F$ fundamental–antifundamental quarks $(Q,\widetilde Q)$, and $W_{\mathrm E}=0$.
Its magnetic description has gauge group $G^d=SU(F-N)$, magnetic quarks $(q,\widetilde q)$, and an elementary gauge singlet $[M]$ corresponding to the electric composite $M=\widetilde QQ$. The magnetic superpotential is \(
 W_{M}=\operatorname{Tr}
 \bigl([M]\widetilde qq\bigr).\) }
  \label{fig:seibergpic}
\end{figure}

Seiberg duality establishes an equivalence between different ultraviolet gauge theory descriptions, which RG flow to the same infrared fixed point~\cite{Seiberg:1994pq}. Figure~\ref{fig:seibergpic} illustrates its original realization in supersymmetric QCD (SQCD).

The original Seiberg duality of SQCD can be extended to adjoint SQCD, yielding similar Seiberg dualities~\cite{Kutasov:1995np,Brodie:1996vx,Kutasov:2014yqa}. For a single $SU(N)$ gauge group with adjoint chirals, duality candidates can be formulated by matching large $N$ superconformal indices as a cyclotomic
polynomial factorization problem~\cite{Bajc:2019vbp}. Finite rank checks and discussions are provided in~\cite{Fang:2024nqy}.

Furthermore, one can generalize Seiberg dualities of single gauge theories to quiver gauge theories. On the one hand, duality can be implemented by a sequence of Seiberg type dualities, dualizing one gauge node at a time~\cite{KlebanovStrassler2000,Feng:2001ToricDuality,BeasleyPlesser2001,FrancoEtAl2006BraneDimers,Xie:2012mr,Franco2012Bipartite,Bao:2020QuiverMutations,KhoLeeSeong2026}. We call these \emph{nodewise dualities}. On the other hand, the duality can be implemented on all the gauge groups at the same time \cite{Brodie:1997sz,Intriligator:1995ax}.  We call these \emph{simultaneous dualities}~\cite{Fang:2026ncq}. 

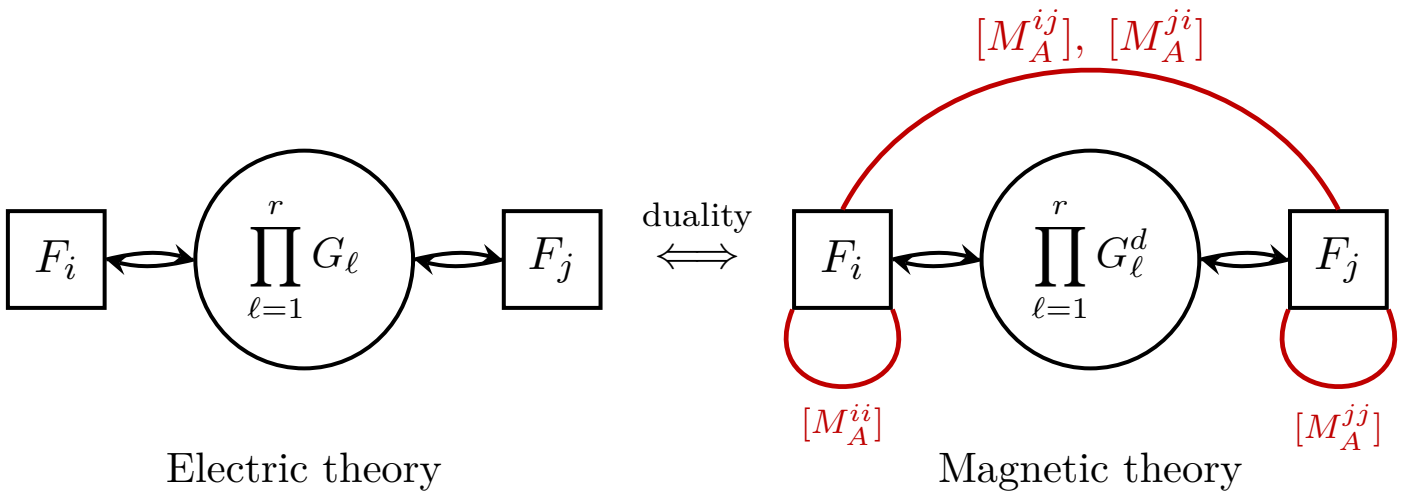
\begin{figure}[H]
  \centering
  \begin{tikzpicture}[
    scale=0.8,transform shape,
    line width=0.75pt, >=stealth,
    gauge/.style={circle,draw,minimum size=18mm,inner sep=1pt,
      align=center,font=\small},
    flavor/.style={rectangle,draw,minimum size=8mm,inner sep=1pt},
    meson/.style={draw=red!75!black,line width=0.9pt}]
    \node[gauge] (EG) at (-3.25,0)
      {$\displaystyle\prod_{\ell=1}^{r}G_\ell$};
    \node[flavor] (EFi) at (-5.30,0) {$F_i$};
    \node[flavor] (EFj) at (-1.20,0) {$F_j$};
    \draw[->] (EFi.east) to[bend left=14] (EG.west);
    \draw[->] (EG.west) to[bend left=14] (EFi.east);
    \draw[->] (EG.east) to[bend left=14] (EFj.west);
    \draw[->] (EFj.west) to[bend left=14] (EG.east);
    \node[font=\small] at (-3.25,-1.75) {Electric theory};

    \node at (0,0) {$\Longleftrightarrow$};
    \node[font=\scriptsize] at (0,0.35) {duality};

    \node[gauge] (MG) at (3.25,0)
      {$\displaystyle\prod_{\ell=1}^{r}G_\ell^d$};
    \node[flavor] (MFi) at (1.20,0) {$F_i$};
    \node[flavor] (MFj) at (5.30,0) {$F_j$};
    \draw[->] (MFi.east) to[bend left=14] (MG.west);
    \draw[->] (MG.west) to[bend left=14] (MFi.east);
    \draw[->] (MG.east) to[bend left=14] (MFj.west);
    \draw[->] (MFj.west) to[bend left=14] (MG.east);
    \draw[meson] (MFi.north)
      .. controls (2.0,1.95) and (4.5,1.95) .. (MFj.north);
    \node[text=red!75!black,font=\small] at (3.25,1.85)
      {$[M_A^{ij}],\ [M_A^{ji}]$};
    \draw[meson] (MFi.south west)
      .. controls +( -0.35,-0.85) and +(0.35,-0.85) .. (MFi.south east)
      node[midway,below,font=\scriptsize,text=red!75!black] {$[M_A^{ii}]$};
    \draw[meson] (MFj.south west)
      .. controls +( -0.35,-0.85) and +(0.35,-0.85) .. (MFj.south east)
      node[midway,below,font=\scriptsize,text=red!75!black] {$[M_A^{jj}]$};
    \node[font=\small] at (3.25,-1.75) {Magnetic theory};
  \end{tikzpicture}
  \caption{Picture of the proposed simultaneous duality. We include adjoints for each gauge group, but omit them from this figure for simplicity.}
  \label{fig:generalpic}
\end{figure}

\begin{figure}[H]
   \centering
  \begin{tikzpicture}[
    line width=0.75pt,
    vertex/.style={circle,draw,fill=white,minimum size=4.4mm,inner sep=0pt},
    type/.style={font=\small},
    scale=0.75,
    transform shape]
    \node[circle,draw,minimum size=22mm,inner sep=1pt,font=\small]
      (ADEproduct) at (-9.05,0.30) {$\displaystyle\prod_{\ell}G_{\ell}$};
    \node[font=\Large] at (-7.53,0.30) {$=$};
    \begin{scope}[shift={(-3.35,2.30)}]
      \node[vertex] (A1) at (-1.50,0) {};
      \node[vertex] (A2) at (-0.75,0) {};
      \node[vertex] (A3) at (0.75,0) {};
      \node[vertex] (A4) at (1.50,0) {};
      \draw (A1) -- (A2);
      \draw (A2) -- (-0.35,0);
      \node at (0,0) {$\cdots$};
      \draw (0.35,0) -- (A3);
      \draw (A3) -- (A4);
      \node[type] at (0,-0.95) {$A_r\quad(r\geq1)$};
    \end{scope}
    \begin{scope}[shift={(3.35,2.30)}]
      \node[vertex] (D1) at (-1.50,0) {};
      \node[vertex] (D2) at (-0.75,0) {};
      \node[vertex] (Dc) at (0.70,0) {};
      \node[vertex] (Du) at (1.50,0.60) {};
      \node[vertex] (Dl) at (1.50,-0.60) {};
      \draw (D1) -- (D2);
      \draw (D2) -- (-0.35,0);
      \node at (0,0) {$\cdots$};
      \draw (0.35,0) -- (Dc);
      \draw (Dc) -- (Du);
      \draw (Dc) -- (Dl);
      \node[type] at (0,-0.95) {$D_r\quad(r\geq4)$};
    \end{scope}
    \begin{scope}[shift={(-3.35,0)}]
      \foreach \i in {0,...,4}
        \node[vertex] (E6-\i) at ({0.70*\i-1.40},0) {};
      \foreach \i [evaluate=\i as \j using int(\i+1)] in {0,...,3}
        \draw (E6-\i) -- (E6-\j);
      \node[vertex] (E6u) at (0,0.65) {};
      \draw (E6-2) -- (E6u);
      \node[type] at (0,-0.60) {$E_6$};
    \end{scope}
    \begin{scope}[shift={(3.35,0)}]
      \foreach \i in {0,...,5}
        \node[vertex] (E7-\i) at ({0.70*\i-1.75},0) {};
      \foreach \i [evaluate=\i as \j using int(\i+1)] in {0,...,4}
        \draw (E7-\i) -- (E7-\j);
      \node[vertex] (E7u) at (-0.35,0.65) {};
      \draw (E7-2) -- (E7u);
      \node[type] at (0,-0.60) {$E_7$};
    \end{scope}
    \begin{scope}[shift={(0,-1.70)}]
      \foreach \i in {0,...,6}
        \node[vertex] (E8-\i) at ({0.70*\i-2.10},0) {};
      \foreach \i [evaluate=\i as \j using int(\i+1)] in {0,...,5}
        \draw (E8-\i) -- (E8-\j);
      \node[vertex] (E8u) at (-0.70,0.65) {};
      \draw (E8-2) -- (E8u);
      \node[type] at (0,-0.60) {$E_8$};
    \end{scope}
  \end{tikzpicture}
  \caption{The product gauge groups $\prod_{\ell}G_{\ell}$ on the left
  can correspond to the finite $ADE$ Dynkin diagrams $\Gamma$ on the right. Each vertex represents a gauge factor
  $G_i=SU(N_i)$, and each edge represents one bifundamental pair.
 Adjoint loops may be present, but they are omitted in this figure for simplicity.}
  \label{fig:ADE-Dynkin}
\end{figure}
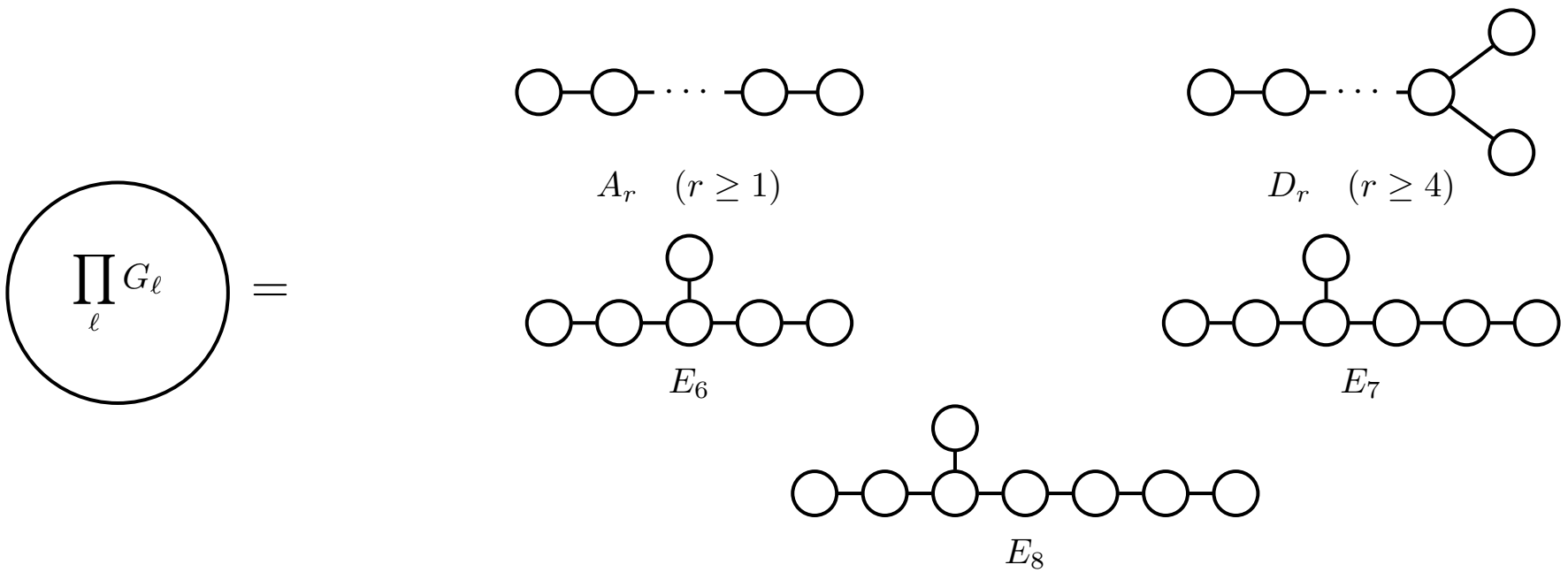

Figure~\ref{fig:generalpic} depicts the proposed simultaneous duality \cite{Fang:2026ncq}. The electric
gauge groups are $\prod_{i=1}^{r}G_i$, with
$G_i=SU(N_i)$ replaced by magnetic gauge groups
$\prod_{i=1}^{r}G_i^d$ and
$G_i^d=SU(N_i^d)$, while the flavor groups are preserved. A square labeled $F_i$ denotes the flavor symmetries, with each one associated with the gauge group $G_i$. The dual theory contains elementary gauge singlets $[M_A^{ij}]$ corresponding to the electric composites
$M_A^{ij}=\widetilde Q_iU_A^{ij}Q_j$. Here $U_A^{ij}$ is an independent composite formed from products of adjoint and bifundamental fields joining vertices $j$ and $i$, and
$A$ labels the meson species.
A huge class of simultaneous duality candidates can be derived from the large-$N$ index strategy, with the 't~Hooft anomaly matching conditions, as well as baryon matching checks provided for finite ranks~\cite{Fang:2026ncq}. This derivation and classification emphasize the information encoded in the theory’s $R$-charges, while in most cases the available tools do not suffice to analyze these candidates; even the superpotential remains undetermined. Accordingly, our goal is to pinpoint at least a few special classes that admit a well-defined superpotential, with an established classical truncation realizing the proposed finite meson spectrum. Therefore, several issues that remain to be resolved in this paper are:
\begin{enumerate}
  \setlength{\itemsep}{1.0pt}
  \setlength{\parskip}{1.0pt}
  \setlength{\parsep}{1.0pt}
  \item Which $R$-charge assignment and quiver graph allow for a finite meson spectrum from the large-$N$ index strategy?
  \item Are there well-defined superpotentials that truncate the mesons, in agreement with the predictions from the large-$N$ index strategy?
  \item Is it possible to realize the simultaneous dualities as a sequence of nodewise dualities?
  \item Is it possible to represent the simultaneous dualities of some gauge quiver theories as a sequence of simultaneous dualities of their subquivers?
\end{enumerate}
With some special $R$-charge assignment, the finite meson requirement selects the gauge quiver represented by the ADE Dynkin diagram depicted in Figure~\ref{fig:ADE-Dynkin}. It also admits a well-defined superpotential that truncates the mesons, in agreement with the predictions from the index. In this specific case, the last two answers are also correct: both a nodewise duality sequence and a subquiver simultaneous duality sequence are possible. Interestingly, the duality proposals and their consistency checks are organized by ADE root systems. The Cartan matrix and Coxeter number enter the magnetic rank map and meson counting, while the Cartan matrix also governs the asymptotically free constraints. Positive roots label the vacuum sectors and transported flavors, and reduced expressions of the longest Weyl-group element organize the nodewise duality sequences.

The paper is organized as follows:
Section~\ref{sec:graph-restrictions} classifies possible gauge graphs and selects the finite ADE quivers. Section~\ref{sec:adeproposal} proposes the simultaneous ADE quiver dualities with a well-defined superpotential,
compares their classical meson species with the index prediction,
and discusses their dynamical conditions.
Section~\ref{sec:nodewise-sequence} uses the ADE Weyl group to describe the rank transformations and construct minimal sequences of nodewise dualities. Section~\ref{sec:subquiver-composition} discusses subquiver simultaneous dualities.
Section~\ref{sec:conclusion} summarizes the results and open questions.

\section{Graph and finite meson restrictions}
\label{sec:graph-restrictions}

To identify which types of quiver gauge graphs and $R$-charge assignments lead to a finite meson spectrum, we begin by revisiting two standard examples of adjoint SQCD, including the Kutasov–Schwimmer's \(A_k\) theory~\cite{Kutasov:1995np} and Brodie’s \(D_{k+2}\) theory~\cite{Brodie:1996vx}.

In Kutasov-Schwimmer's $A_k$
theory~\cite{Kutasov:1995np}, the superpotential and $R$-charge of the adjoint chiral field are
\begin{equation}
 W_{A_k}=\operatorname{Tr}X^{k+1},
 \qquad R(X)=\frac{2}{k+1}.
\end{equation}
The full-matrix F-term equation $X^k=0$ reduces the adjoint composite to $1,X,\ldots,X^{k-1}$. The resulting mesons are
\begin{equation}
 M_a=\widetilde QX^aQ,\qquad 0\leq a\leq k-1.
\end{equation}

In Brodie's $D_{k+2}$ theory~\cite{Brodie:1996vx}, the superpotential and $R$-charges of the adjoint chiral fields are
\begin{equation}
 W_{D_{k+2}}
 =\operatorname{Tr}\left(\frac{X^{k+1}}{k+1}+XY^2\right),
 \qquad R(X)=\frac{2}{(k+1)},\, R(Y)=\frac{k}{(k+1)}.
\end{equation}
Here the mixed interaction constrains two noncommuting adjoints:\begin{equation}X^k+Y^2=0,\qquad XY+YX=0.\end{equation}
For \emph{odd} $k$, these full-matrix equations imply $Y^3=0$, the resulting undressed mesons are  $$X^aY^b,\qquad 0\leq a<k,\,0\leq b\leq2.$$ 
For even $k$, these classical relations do not establish the same truncation. Therefore, this case requires additional quantum relations arising from the duality requirement~\cite{Brodie:1996vx}. Other similar examples include the $E_7$ theory~\cite{Kutasov:2014yqa}, and the
$M_k$ and $N_k$ theories~\cite{Bajc:2019vbp}. None of these theories has a well-defined superpotential that allows a classical truncation producing the proposed finite meson spectrum; instead, quantum chiral-ring relations are imposed. \footnote{Although we concentrate on dualities with a well-defined superpotential in this paper, this should not be taken as a criterion for excluding potential dual candidates.}

Thus, when moving from a single gauge group to a gauge quiver, the problem becomes identifying which gauge
graphs and adjoint matter assignments allow for a finite collection of
independent meson types with the specified $R$-charges.
We use the large-$N$ index strategy to derive the simultaneous duality~\cite{Fang:2026ncq} and find the gauge graphs and adjoint assignments satisfying it.

Let $\Gamma$ be a connected graph with $r$ gauge
vertices. Vertices correspond to the gauge group $\mathcal G=\prod_iSU(N_i)$, and each edge between vertices corresponds to one
bifundamental pair $X_{ij},X_{ji}$. We assume that each gauge group is accompanied by an ordinary flavor group with $F_i>0$, and that an adjoint field $X_i$ may or may not be present. The proposed magnetic theory has the same gauge node
graph and adjoint assignment as the electric theory. The electric meson can be written as
$M_A^{ij}=\widetilde Q_iU_A^{ij}Q_j$.
Here the undressed meson $U_A^{ij}$ consists of a product of adjoint and bifundamental fields along a single open path, with $A$ distinguishing the meson species. We denote its $R$-charge by
$I_A^{ij}=R(U_A^{ij})$, so that the (dressed) meson $M_A^{ij}$ carries $R$-charge $R(\widetilde Q_i)+R(Q_j)+R(U_A^{ij})$. To characterize the finite mesons, we briefly review several definitions and the large-$N$ index strategy used in \cite{Fang:2026ncq}, as follows.

The matrix $M_Q(t)$ records the $R$-charges of the chiral fields of
the gauge quiver $Q$: its $(i,j)$ entry is the sum of $t^{R(X_{ij})}$ over
fields $X_{ij}$ with gauge indices from vertex $i$ to vertex $j$. Therefore,  bifundamental fields
contribute off-diagonal entries \(\sum t^{R(X_{ij})}\) and adjoints contribute diagonal entries \(\sum t^{R(X_{ii})}\). Ordinary quarks supply the
external flavor indices of the mesons and are not included in this matrix $M_Q(t)$.

The quiver Hilbert series entering the large-$N$ index strategy is defined by \cite{Bocklandt2008CY3,Eager2014AMaxVolume} \footnote{Here $H(Q,t)$ is the rational matrix determined by the quiver
matter content and its assigned $R$-charges. It need not equal
the matrix Hilbert series of the classical F-term algebra in general.
For the full-matrix ADE superpotential introduced in
Section~\ref{sec:adeproposal}, we prove that the classical
meson-counting polynomial agrees with the index prediction in Appendix \ref{app:meson-graded-matching}.
\begin{equation}
\operatorname{Hilb}(\mathcal B_k;y)
=(1-y^{2kh})H(Q,y^{k+1})
=\mathcal P_{2kh}(y).
\end{equation}}

\begin{equation}\label{eq:graph-index-kernel-definition}
 H(Q,t)=\bigl[(1-t^2)I_r-M_Q(t)+t^2M_Q(t^{-1})^T\bigr]^{-1}.
\end{equation}
Here $I_r$ is the identity matrix. Matching the electric and
magnetic superconformal indices in the large-$N$ limit under the duality ansatz gives~\cite{Fang:2026ncq}
\begin{equation}\label{eq:ade-phys-kernel}
 \begin{aligned}
 \mathcal P_L(y)
 &=H(Q,y^\ell)-y^LH(Q,y^\ell)^T\\
 &\xrightarrow{H=H^T}(1-y^L)H(Q,y^{k+1}),\qquad
 L=\ell(\Delta+2)\in\mathbb Z_{>0}.
 \end{aligned}
\end{equation}
Because we focus on situations in which bifundamental fields connecting gauge groups carry identical $R$-charges, the second line is a direct consequence of the symmetry between the bifundamental pairs. Here $\ell$ is chosen such that every exponent in $H(Q, t)$ becomes an integer after setting
$t= y^\ell$.
The parameter $\Delta$ specifies the pairing of complementary
$R$-charges, because in our duality ansatz the mesons are expected to be paired.\begin{equation}I_A^{ij}+I_{A'}^{ij}=\Delta.\end{equation}

With $t=y^\ell$,
the index equation predicts the matrix equation
\begin{equation}\label{eq:ade-phys-mesons}
 (\mathcal P_L(y))_{ij}
 =\sum_A y^{\ell I_A^{ij}}.
\end{equation}
 Each coefficient counts species at a fixed undressed $R$-charge. Thus, finitely
many electric meson composites (or elementary magnetic singlets) require every entry of
$\mathcal P_L$ to be a finite polynomial with nonnegative integer
coefficients. The total number of species in channel $(i,j)$ is
$a_{ij}=(\mathcal P_L(1))_{ij}$.

 Guided by insights from the $A_k$ and $D_{k+2}$ theories, we consider the most possible $R$-charge assignments with a finite meson spectrum. Consider two cases: (i) a quiver that includes an adjoint on each gauge node with $R(X_i)=\frac{2}{(k+1)}$, together with bifundamental fields connecting certain pairs of gauge nodes with $R(X_{ij})=\frac{k}{(k+1)}$; and (ii) a quiver with no adjoint fields, where bifundamentals between some selected gauge nodes have $R(X_{ij})=\frac{1}{(k+1)}$. Given these assumed $R$-charge assignments, the gauge quiver graph structure will be determined separately in later sections \ref{sec:full-adjoint-criterion} and \ref{sec:withouadjointdualcriterion}.

\subsection{Quivers with adjoints: $R(X_i)=\frac{2}{(k+1)}$ and
$R(X_{ij})=\frac{k}{(k+1)}$}
\label{sec:full-adjoint-criterion}

We start by specifying the setup: for any pair of gauge groups $N_i$ and $N_j$, there may or may not exist a pair of bifundamental fields $X_{ij},X_{ji}$. If these fields are present, they carry the same $R$-charge $\frac{k}{k+1}$. In addition, each gauge node always contains an adjoint field with $R$-charge $\frac{2}{k+1}$, namely
\begin{equation}\label{eq:ade-phys-charges}
 R(X_i)=\frac{2}{k+1},\qquad
 R(X_{ij})=R(X_{ji})=\frac{k}{k+1}.
\end{equation}
Let $G_\Gamma$ denote the adjacency matrix of $\Gamma$, where $G_{ij}=1$ if the vertices $i$ and $j$ are adjacent (i.e., $i\sim j$), and $G_{ij}=0$ otherwise. Then the matrix \(M_Q\) is given by
\begin{equation}
 M_Q(t)=t^{\frac{2}{(k+1)}}I_r
          +t^\frac{k}{(k+1)}G_\Gamma,
          \end{equation}
Take \(\ell = k+1\) and substitute $t = y^{\ell}$, then the quiver Hilbert series \eqref{eq:graph-index-kernel-definition} becomes
 \begin{equation}  \label{eq:adjoint-index-kernel}      H(Q,y^{k+1})=\frac{1}{1-y^2}
 \left[(1+y^{2k})I_r-y^kG_\Gamma
       \right]^{-1}.
\end{equation}
 
Substitute the adjacency matrix by the matrix $C$, defined as $C=2I_r-G_\Gamma$, which yields
\begin{equation}
 H(Q,y^{k+1})
 =\frac{1}{1-y^2}
 \bigl[(1-y^k)^2I_r+y^kC\bigr]^{-1}.
\end{equation}
Under this framework, we state the following theorem to determine the gauge graph:

\paragraph{Theorem (ADE classification with adjoints at every node)}
\textit{Consider a quiver whose gauge
graph $\Gamma$ is finite, connected, and simple\footnote{The adjoints \(X_i\) contributions are specified separately and are not counted as loops in \(\Gamma\).}, with $r$ vertices,
an adjoint and ordinary flavor at every gauge node, and the
$R$-charges specified in \eqref{eq:ade-phys-charges}.
The finite-meson requirement is satisfied if and only if $\Gamma$ is a finite ADE Dynkin diagram:
}
\[
\boxed{
\left(
\exists L\in\mathbb Z_{>0}:\quad
\mathcal P_L(y)\in\mathbb Z_{\geq0}[y]^{r\times r}
\right)
\quad\Longleftrightarrow\quad
\Gamma\text{ is a finite ADE Dynkin diagram}.
}
\]
\textit{For an ADE diagram with Coxeter number $h$,
one admissible choice is $L=2kh$.}
 
To see why ADE is necessary, recall that $C=2I_r-G_\Gamma$ is
real and symmetric. For an eigenvector $v\ne0$ with $Cv=cv$, the
meson-counting matrix \eqref{eq:ade-phys-kernel} acts as
\begin{equation}
 \mathcal P_L(y)v
 =\frac{1-y^L}
 {(1-y^2)\bigl[(1-y^k)^2+cy^k\bigr]}\,v.
\end{equation}
A finite meson spectrum requires this scalar function to be a
polynomial. If $c<0$, the factor
$d_c(y)=(1-y^k)^2+cy^k$ has $d_c(0)=1$ and $d_c(1)=c<0$.
It therefore vanishes at some $y_0\in(0,1)$, where
$1-y_0^L>0$, leaving an uncancelled pole. If $c=0$, the
denominator has a third-order zero at $y=1$, while the numerator
has only a simple zero, so a pole again remains. Every
eigenvalue of $C$ must consequently be positive:
\begin{equation}\label{eq:C>0}
 \boxed{\mathcal P_L(y)\text{ polynomial}
 \ \Longrightarrow\ C\succ0.}
\end{equation}
Here $C\succ0$ means positive definiteness. For a finite
connected simple graph, the classification of positive-definite
simply laced Cartan matrices gives~\cite{Dynkin1947,EtingofLie}
\begin{equation}\label{eq:ADE}
 \boxed{C\succ0\quad\Longleftrightarrow\quad
 \Gamma\text{ is a finite ADE Dynkin diagram}.}
\end{equation}
Now we have proved that the finite meson requires the graph to be the ADE Dynkin diagram. Conversely, sufficiency is established by setting $L = 2kh$ and is discussed in section \ref{sec:adeproposal}.

\subsection{Quivers without adjoints: $R(X_{ij})=\frac{1}{(k+1)}$}\label{sec:withouadjointdualcriterion}

One can instead consider a gauge quiver without adjoint fields: each edge connecting two chosen gauge nodes $N_i$ and $N_j$ represents a pair of bifundamental chiral fields $X_{ij}$ and $X_{ji}$, carrying the same $R$-charge
\begin{equation}
 R(X_{ij})=R(X_{ji})=\frac{1}{k+1}.
\end{equation}
In this setting, we have
$M_Q(t)=t^{\frac{1}{(k+1)}}G_\Gamma$. Plugging this into
\eqref{eq:graph-index-kernel-definition} and then substituting $t=y^{\ell}$ with $\ell=k+1$ yields the matrix Hilbert series
\begin{equation}\label{eq:no-adjoint-index-kernel}
 H(Q,y^{k+1})=\frac{1}{1-y^2}\left[
       \left(\sum_{a=0}^{k}y^{2a}\right)I_r
       -y\left(\sum_{a=0}^{k-1}y^{2a}\right)G_\Gamma\right]^{-1}.
\end{equation}

When $k=1$, this reduces to a special case of the ADE proposal with adjoints at every node, as discussed in Section \ref{sec:full-adjoint-criterion}. As $k$ grows, this charge assignment becomes restricted to linear quivers $A_3$ and $A_2$, as summarized in Table~\ref{tab:non-adj-class}. The detailed proof of Table~\ref{tab:non-adj-class} is provided in
Appendix~\ref{app:non-adj-proof}.
 
\begin{table}[H]
 \centering
 \begin{tabular}{c c}
 \toprule
 $k$ & Allowed gauge graphs\\
 \midrule
 $1$ & Any finite ADE Dynkin diagram\\
 $2$ & $A_2$ or $A_3$\\
 $\geq3$ & $A_2$\\
 \bottomrule
 \end{tabular}
 \caption{A complete classification of quivers without adjoint satisfying the finite-meson requirement for \(R(X_{ij})=\frac{1}{(k+1)}\), considering gauge graphs with at least two vertices and standard flavors at each gauge node.}
 \label{tab:non-adj-class}
\end{table}

After determining the $R$-charges and the gauge graphs so that the independent mesons remain finite in this section, we next consider a well-defined superpotential that establishes classical truncation realizing the proposed finite meson spectrum with adjoints at every node case for $k\geq 1$ and present the ADE duality proposals. As for the quiver without adjoint case,  the $A_2$ family of dualities for $k\geq 1$ has been discussed in~\cite{Intriligator:1995ax}, and its nodewise duality realization is discussed in \cite{Fang:2026ddf}. The $A_3$ entry at $k=2$ qualifies as a candidate under the large-$N$ index strategy, but no superpotential realization has been identified, and establishing a physical duality in this case requires a separate analysis.

\section{ADE duality proposals}	\label{sec:adeproposal}
The preceding section shows that with an adjoint at every gauge node, the finite meson requirement selects finite ADE Dynkin diagrams. We now construct a superpotential whose classical F-term relations yield the meson spectrum corresponding to $L=2kh$, where $h$ is the Coxeter number, and present the associated magnetic duality proposals. This choice of $L$ is determined by
the superpotential realization; the graph discussions alone require only the existence of an admissible $L$.

\subsection{The electric theory}

We consider an electric gauge theory described by a quiver whose Dynkin diagram is
$\Gamma=A_r,D_r,E_6,E_7,E_8$. At each node, there is an $SU(N_i)$ gauge group together with an adjoint chiral multiplet. In addition to the gauge quiver, each gauge factor $SU(N_i)$ is coupled to $F_i$ fundamental–antifundamental pairs $(Q_i,\widetilde Q_i)$. We combine the traceless adjoint component and the trace singlet into a single full-matrix field
\begin{equation}
 X_i=\widehat X_i+x_iI_{N_i},\qquad
 \operatorname{Tr}\widehat X_i=0,\qquad
 x_i=\frac{1}{N_i}\operatorname{Tr}X_i.
\end{equation}
Hence, $X_i$ transforms in $\mathrm{Adj}_{SU(N_i)}\oplus\mathbf1$ and is chosen in this paper.~\footnote{In this paper, we focus on dualities with traceful field $X_i$. One may alternatively study dualities with a traceless adjoint $\hat X_i$. Traceful or traceless electric theories share the same independent meson generators in the large-$N$ limit discussed in Appendix \ref{app:meson-basis-completeness}. Both traceful and traceless dualities can likewise be formulated as well-defined finite-rank dualities, because traceful and traceless forms of the elementary Kutasov--Schwimmer duality are established~\cite{Bajeot:2023gyl}, as long as all induced interactions and necessary Higgs reductions are handled consistently within the nodewise realization.}As specified in Section \ref{sec:graph-restrictions}, the $R$-charges are:
\begin{equation}\label{eq:ade-spectrum}
 R(X_i)=\frac{2}{k+1},\qquad
 R(X_{ij})=R(X_{ji})=\frac{k}{k+1}.
\end{equation}
For $k=1$, the adjoint fields are massive and can therefore be integrated out since $R(X_i)=1$, which corresponds to the case shown in the first row of Table \ref{tab:non-adj-class}.

\paragraph{Superpotential and F-term equations}
Let $\mathcal O$ contain one ordered pair $(i,j)$ of each edge, the superpotential of the electric theory is~\cite{CachazoEtAl2002,CachazoKatzVafa2001}
\begin{equation}\label{eq:ade-W-electric}
 \boxed{\displaystyle
 W_E=\frac{1}{k+1}\sum_i\operatorname{Tr}X_i^{k+1}
 +\sum_{(i,j)\in\mathcal O}
 \left[\operatorname{Tr}(X_iX_{ij}X_{ji})
       -\operatorname{Tr}(X_jX_{ji}X_{ij})\right].}
\end{equation}

The F-term relations are
\begin{equation}\label{eq:ade-full-terms}
 X_iX_{ij}=X_{ij}X_j,\qquad
 \mu_i+X_i^k=0,\qquad 
 \text{with }\mu_i=\sum_{(i,j)\in\mathcal O}X_{ij}X_{ji}
       -\sum_{(j,i)\in\mathcal O}X_{ij}X_{ji}.
\end{equation}
Preserving the trace singlets gives the same matrix F-term
equations as in the unitary quiver
constructions~\cite{CachazoEtAl2002,CachazoKatzVafa2001}.
The relation $X_iX_{ij}=X_{ij}X_j$ allows adjoint insertions
to be moved along a bifundamental path and collected at
one endpoint. Consequently, every meson can be written
as a linear combination of operators of the form
\begin{equation}
 \widetilde Q_iX_i^s
 X_{ii_1}X_{i_1i_2}\cdots X_{i_{m-1}j}Q_j.
\end{equation}
The undressed mesons $X_iX_{ii_1}X_{i_1i_2}\cdots X_{i_{m-1}j}$ can be separated into two parts: one is an adjoint composite $X_i^s$, and the other is a bifundamental composite. The next step is to construct a complete basis modulo F-term relations.

\paragraph{Independent meson basis}
For ordered endpoints $(i,j)$, let $U_A^{ij}(X)$ range over a basis of independent gauge field combinations modulo the F-term relations. Then by attaching the $\widetilde Q_i$ and $Q_j$ fields, one obtains the corresponding set of classical meson species as
\begin{equation}
 \boxed{M_A^{ij}=\widetilde Q_iU_A^{ij}(X)Q_j,\qquad
 R(M_A^{ij})=R(\widetilde Q_i)+R(Q_j)+R(U_A^{ij}).}
\end{equation}
Thus $A$ labels a species, and each species has $F_iF_j$ flavor
components. The flavor indices transform as those of
$\widetilde Q_i$ and $Q_j$.

 Let $G$ be the
adjacency matrix. Define
$\nu_m=(\nu_m^{ij})$ to count the independent
starting paths containing $m$ bifundamentals from node $j$
to node $i$~\cite{etingof2006centralextensionspreprojectivealgebras}: \footnote{Using the notation $P_m(G)$ from~\cite{etingof2006centralextensionspreprojectivealgebras}, we have $\nu_m=P_m(G)=P_m(2I_r-C)$.}
\begin{equation}
 \nu_0=I_r,\qquad \nu_1=G,\qquad
 \nu_{m+1}=G\nu_m-\nu_{m-1},\qquad 1\leq m \leq h-2,
\end{equation}
where $I_r$ is the $r\times r$ identity matrix. For $m=0$, the starting combination is the identity at $i=j$;
for $m=1$, it is $X_{ij}$ when the edge exists. Larger $m$ corresponds to product terms of the form $X_{i i_1} X_{i_1 i_2}\cdots X_{i_{m-1}j}$. 

The meson basis is summarized as follows:
\begin{equation}\label{eq:meson-spectrum}
\boxed{
M_{m,\alpha,s}^{ij}
=
\widetilde Q_iX_i^sB_{m,\alpha}^{ij}Q_j,
\qquad
\begin{gathered}
1\leq i,j\leq r,\qquad 0\leq m\leq h-2,\\
1\leq\alpha\leq\nu_m^{ij},\qquad
0\leq s\leq k(h-1-m)-1.
\end{gathered}
}
\end{equation} These general results recover known results: $A_2$
dualities~\cite{Brodie:1996vx} for any $k$ and $A_3$~\cite{Brodie:1997sz} for odd $k$. For any integer $k\geq1$ and any finite ADE gauge graph,
Appendix~\ref{app:meson-basis-completeness} derives the
basis~\eqref{eq:meson-spectrum} from the F-term algebra in detail,
using the results of~\cite{etingof2006centralextensionspreprojectivealgebras}. We choose each representative $B_{m,\alpha}^{ij}$
to be a single product of bifundamentals.\footnote{Other
homogeneous bases may use linear combinations of paths;
the graded meson counts and $R$-charges remain unchanged.}
Each starting path admits $k(h-1-m)$ adjoint
contributions, with $0\leq s\leq k(h-1-m)-1$. Together, these
paths form a complete, linearly independent basis modulo
the F-term relations.

Since $U_A^{ij}=X_i^sB_{m,\alpha}^{ij}$, its $R$-charge is
\begin{equation}
 R(U_A^{ij})=\frac{km+2s}{k+1}.
\end{equation}
In Appendix \ref{app:meson-graded-matching}, we also prove that the mesons are directly derived from the F-term algebra
\begin{equation}\label{eq:ade-meson-graded-match}
 \sum_A y^{(k+1)R(U_A^{ij})}=\bigl(\mathcal P_{2kh}(y)\bigr)_{ij}.
\end{equation}
 
matches the prediction from the large-$N$ strategy \eqref{eq:ade-phys-kernel}.
 
\subsection{Seiberg duality proposal}
This subsection proposes the simultaneous duality of ADE quiver gauge theories by applying the large-$N$ index approach~\cite{Fang:2026ncq}. Let $\Gamma$ be a finite ADE Dynkin diagram with $r$ vertices, adjacency matrix $G$, Cartan matrix $C=2I_r-G$, and Coxeter number $h$. The electric and proposed magnetic gauge groups are
\begin{equation}
 \prod_{i=1}^{r}SU(N_i),
\quad\text{and}\quad
\prod_{i=1}^{r}SU( N_i^d).
\end{equation}
The magnetic matrix field $Y_i$ likewise contains a traceless
adjoint of $SU(N_i^d)$ and an independent trace singlet.
 The flavor groups $\prod_i SU(F_i)_L\times SU(F_i)_R$ are unchanged.
For the $L=2kh$ realization selected above, the rank ansatz gives
\begin{equation}\label{eq:dual-rank}
a_{ij}:=kh(C^{-1})_{ij},
\qquad
  N_i^d=\sum_{j=1}^{r}a_{ij}F_j-N_i.
\end{equation}
Express the dual gauge group ranks in matrix form as:
\begin{equation} \label{eq:dual-ranks}
\boldsymbol{N^d}=khC^{-1}\boldsymbol F-\boldsymbol N.
\end{equation}

The $R$-charges of the classical undressed mesons are
\begin{equation}
   \{ I^{ij}\}=\{\frac{km+2s}{k+1}\},\text{ with } 0\leq m\leq h-2,\quad 0\leq s\leq k(h-1-m)-1.
\end{equation}
  
For each $(m,s)$ there are $\nu_m^{ij}$ species. The multiplicity
at a fixed charge is obtained by summing over all pairs $(m,s)$ with that charge.
The complementary undressed $R$-charges entering the magnetic
meson couplings sum to
\begin{equation}\label{eq:ade-meson-pairing}
 \Delta=\frac{2kh}{k+1}-2.
\end{equation}

Following the conventions of~\cite{Fang:2026ncq}, we summarize
the matter content, representations, and $R$-charges of the
electric theory and its proposed magnetic dual in
Table~\ref{tab:matter-fields}.

\begin{table}[H]
   \centering
   \resizebox{\textwidth}{!}{%
    \begin{tabular}{|ccccccc|}
        \hline
        Electric field & Representation & $R$-charge && Magnetic field & Representation & $R$-charge\\ \hline
        $Q_i$ & $(\mathbf N_i,\mathbf F_{i,L})$
       &$R(Q_i)=1-\frac{(C\boldsymbol N)_i}{(k+1)F_i}$ & $\longleftrightarrow$ &
        $\tilde q_i$ & $(\overline{\mathbf N}_i^d,\overline{\mathbf F}_{i,L})$ &$R(\widetilde q_i)=1-\frac{(C \boldsymbol N^d)_i}{(k+1)F_i}$\\ \hline
       
        $\tilde Q_i$ & $(\overline{\mathbf N}_i,\mathbf F_{i,R})$&$R(\widetilde Q_i)=1-\frac{(C\boldsymbol N)_i}{(k+1)F_i}$
        & $\longleftrightarrow$ & $q_i$ & $(\mathbf N_i^d,\overline{\mathbf F}_{i,R})$ &$R(q_i)=1-\frac{(C \boldsymbol N^d)_i}{(k+1)F_i}$\\ \hline
        
        $X_{ij}$ & $(\mathbf N_i,\overline{\mathbf N}_j)$
        &$\frac{k}{k+1}$& $\longleftrightarrow$ &$Y_{ji}$ & $(\overline{\mathbf N}_i^d,\mathbf N_j^d)$&$\frac{k}{k+1}$ \\ \hline
           $X_i$ & $\mathrm{Adj}_i\oplus\mathbf1$
        &$\frac{2}{k+1}$& $\longleftrightarrow$ &$Y_i$ & $\mathrm{Adj}_i^d\oplus\mathbf1$&$\frac{2}{k+1}$ \\ \hline
       $M_A^{ij}$ &$(\mathbf F_{i,R},\mathbf F_{j,L})$&$R(\widetilde Q_i)+R(Q_j)+I_A^{ij}$&$\longleftrightarrow$& $[M_A^{ij}]$&$(\mathbf F_{i,R},\mathbf F_{j,L})$&$R(\widetilde Q_i)+R(Q_j)+I_A^{ij}$\\ \hline
    \end{tabular}
   }
    \caption{Matter fields and mesons in the duality proposal. The labels $L,R$ identify the flavor factors, and all unlisted factors act trivially under $\prod_{i=1}^r SU(N_i)SU(F_i)_L SU(F_i)_R$. For example, $(\mathbf N_i,\mathbf F_{i,L})$ denotes a fundamental of $SU(N_i)$ and $SU(F_i)_L$, $\mathrm{Adj}_i$ denotes the adjoint representation of $SU(N_i)$, the additional $\mathbf1$ is the trace singlet. }
    \label{tab:matter-fields}
\end{table}
In the electric theory, $M_A^{ij}$ is a composite chiral operator.
In the dual magnetic theory, $[M_A^{ij}]$ denotes an elementary
chiral singlet with the same flavor representation and $R$-charge. 
Introduce magnetic composites
\begin{equation}
 M_{m,A}^{ji}=\widetilde q_j\mathcal V_A^{ji}(Y)q_i,
 \qquad R(\mathcal V_A^{ji})=\Delta-I_A^{ij},
 \qquad I_A^{ij}=R(U_A^{ij}).
\end{equation}
Here $\mathcal V_A^{ji}(Y)$ is a product of
magnetic adjoints $Y_i$ and bifundamentals $Y_{ij}$ with reversed endpoints and
complementary undressed $R$-charge. 
The magnetic superpotential is
\begin{equation}\boxed{
\begin{aligned}
W_M&=\frac{1}{k+1}\sum_i\operatorname{Tr}Y_i^{k+1}+\sum_{(i,j)\in\mathcal O}
 \left[\operatorname{Tr}(Y_iY_{ij}Y_{ji})
       -\operatorname{Tr}(Y_jY_{ji}Y_{ij})\right]\\
&\quad+\sum_{i,j,A} [M_A^{ij}]M_{m,A}^{ji}.
\end{aligned}}
\end{equation}

\subsection{Dynamical consistency}
\label{sec:dynamical-consistency}

To complement the meson spectrum matching, we now examine two dynamical consistency checks of the proposed dual pair: simultaneous one-loop asymptotic freedom and the existence of supersymmetric vacua.

\paragraph{Simultaneous asymptotic freedom}
For $k\geq2$, under the proposed adjoint and bifundamental matter pattern, $F_i$ flavor pairs are attached to gauge node $i$. 

For each electric or magnetic gauge coupling $g_i$, the beta function is given by \cite{Novikov1986Beta,Barnes:2005}
\[
 \boxed{\beta_{g_i}:=\mu\frac{dg_i}{d\mu}
 =-\frac{g_i^3}{16\pi^2}\,b_i+\text{higher-loop terms},}
\]
where $b_i$ is the one-loop coefficient. Thus $b_i>0$ means that the gauge
factor is asymptotically free at one loop.

With
$C=2I_r-G$ and the proposed magnetic ranks
$\boldsymbol N^d=khC^{-1}\boldsymbol F-\boldsymbol N$, the electric
and magnetic one-loop coefficients of the beta functions are
\begin{equation}
 \boldsymbol b^{\mathrm E}=C\boldsymbol N-\boldsymbol F,
 \qquad
 \boldsymbol b^{\mathrm M}=(kh-1)\boldsymbol F-C\boldsymbol N.
\end{equation}

Both the electric and magnetic theories have all gauge factors that are asymptotically free provided that $\boldsymbol b^{\mathrm E}>0$ and $\boldsymbol b^{\mathrm M}>0$, namely,
\begin{equation}\label{eq:dyn-af}
 \boxed{\boldsymbol F<C\boldsymbol N<(kh-1)\boldsymbol F.}
\end{equation}

For $k=1$, the adjoints are massive and must first be integrated out. Consequently, the asymptotically free region should be treated separately. Writing $D=C+I_r=3I_r-G$, the coefficients of the 1-loop beta functions are
\begin{equation}\boldsymbol b^{\mathrm E}=D\boldsymbol N-\boldsymbol F,\qquad \boldsymbol b^{\mathrm M}=D\boldsymbol N^d-\boldsymbol F,\end{equation}
with $\boldsymbol N^d=hC^{-1}\boldsymbol F-\boldsymbol N$. Therefore, the asymptotically  free region when $k=1$ is 
\begin{equation}
 \boxed{\boldsymbol F<D\boldsymbol N<
 \left[h(I_r+C^{-1})-I_r\right]\boldsymbol F.}
\end{equation}
For the $A_1$ at $k=1$, the asymptotically free region is $\tfrac32N<F<3N$, reproducing the conformal window of SQCD~\cite{Seiberg:1994pq}.

\paragraph{Stable vacua.}
Recall that $SU(N)$ SQCD with $F$ flavors admits supersymmetric vacua for $F\geq N$. For $F<N$, a dynamically
generated superpotential produces a runaway~\cite{Seiberg:1994pq}. For Kutasov--Schwimmer's $A_k$ theory~\cite{Kutasov:1995np,Kutasov:1995ss}, we add generic lower-degree
terms to the adjoint polynomial $W(z)$ while retaining its
leading term $z^{k+1}$. For generic coefficients, the adjoint polynomial $W(z)$ has
$k$ distinct critical points $z_l$, defined by
$W'(z_l)=0$. Let $n_l\in\mathbb Z_{\geq 0}$ be the number of colors assigned to the $l$-th critical point, so that $N=\sum_{l=1}^k n_l$.
The adjoint becomes massive on these generic branches, leaving each
nonabelian $SU(n_l)$ SQCD sector with $F$ flavors. Avoiding a runaway requires $F\geq n_l$ separately. Adding these terms together and taking the quotient by $k$, the stability region is given by\cite{Kutasov:1995np} \[F\geq \frac{N}{k}.\] Since the highest-degree term is retained, they argue
that the existence of a vacuum is preserved when the deformation is removed. We apply the same deformation and argument to our ADE gauge quivers.

We deform the electric
superpotential by adding small, generic lower-degree terms. Let $X_i$ denote the adjoint matrix, and let $z\in\mathbb C$ be an eigenvalue. We write the purely adjoint contribution to the superpotential \eqref{eq:ade-W-electric} as
\begin{equation}
W_i(z)=\frac{g_i}{k+1}z^{k+1}+\text{terms of lower degree}.
\end{equation}

For a general positive root \(\alpha=\sum_i\alpha_i\boldsymbol e_i\), the block contains \(\alpha_i\) color components at node \(i\). Summing the F-term equations on the block cancels every bifundamental contribution and gives~\cite{CachazoEtAl2002}
\begin{equation}
 W_\alpha'(z)=\sum_i\alpha_iW_i'(z)=0.
\end{equation}
  Here $z$ is the common eigenvalue on the root blocks. The adjoint-bifundamental interactions lead to the weighted condition $ W_\alpha'(z)=0$, rather than separate conditions $ W_i'(z)=0$. 
If $\sum_i\alpha_i g_i\ne0$ for every positive root, generic lower
coefficients give $k$ distinct solutions for each $\alpha$. Let $M_{\alpha,\ell}\in\mathbb Z_{\ge0}$ count blocks associated with
root $\alpha$ at its $\ell$th critical point, and write
$\boldsymbol\alpha=(\alpha_1,\ldots,\alpha_r)^T$. Their contributions to the color ranks and the number of flavors obey
\begin{equation}
N_i=\sum_{\alpha>0}\sum_{\ell=1}^{k}\alpha_iM_{\alpha,\ell},
 \qquad F_\alpha=\sum_i\alpha_iF_i.
\end{equation}
The factor $\alpha_i$ appears in both expressions because the root
block has that many color components at node $i$. For a nontrivial
low-energy SQCD factor with $F_\alpha$ flavors, the familiar stable vacuum region gives
$  F_\alpha\geq M_{\alpha,\ell}$~\cite{Seiberg:1994pq}. Summing over the $k$ critical points and dividing by an overall $k$ gives the stable vacuum region:
\begin{equation}\label{eq:stable}
 \boxed{\boldsymbol N=\sum_{\alpha>0}m_\alpha\boldsymbol\alpha,
 \qquad m_\alpha\in\mathbb Z_{\ge0},\qquad F_\alpha\geq \frac{m_\alpha}{k},
 \qquad m_\alpha=\sum_{\ell=1}^{k}M_{\alpha,\ell}.}
\end{equation}
For $A_1$, it reduces to
$N\leq kF$, the familiar bound in Kutasov--Schwimmer duality
\cite{Kutasov:1995np,Kutasov:1995ss}.

\section{Nodewise sequences and ADE rank transformations}
\label{sec:nodewise-sequence}
The two-node dualities \cite{Intriligator:1995ax} listed in Table \ref{tab:non-adj-class} can be reproduced by sequences of Seiberg duality~\cite{Seiberg:1994pq} and Kutasov--Schwimmer duality \cite{Kutasov:1995np}, together with Higgsing and integrating out massive fields, to reproduce the simultaneous proposal in~\cite{Fang:2026ddf}. The corresponding chiral spectra and superpotentials are presented step by step in~\cite{Fang:2026ddf}.

In this section, we extend the nodewise construction from the particular two-node example to the general ADE simultaneous duality proposals discussed in Section~\ref{sec:adeproposal}, with superpotential \eqref{eq:ade-W-electric}. A detailed discussion of the intermediate superpotentials is omitted, as it closely parallels the two-node case. We use the positive root and Weyl reflections to describe the intermediate gauge ranks along the sequence and nodewise orderings.

Each nodewise duality step is performed using the full inherited
superpotential. For $k=1$, it consists of Seiberg duality followed
by integrating out massive fields. For $k\geq2$, it consists of
Kutasov--Schwimmer (KS) duality, the induced Higgsing and
integrating out massive fields:
\begin{equation}\label{eq:nodewise-step}
\begin{aligned}
& k=1\text{ step}:\quad \text{Seiberg duality}
             \longrightarrow\text{Integrating out massive fields},\\
& k\ge2\text{ step}:\quad \text{KS duality}
             \longrightarrow\text{Higgsing}
             \longrightarrow\text{Integrating out massive fields}.
\end{aligned}\end{equation}
For fixed $k$, our proposed realization uses the corresponding type of step successively on different gauge nodes throughout the sequence. This differs from the construction in~\cite{Fang:2026ddf}, which employs both types of steps $k\geq 2$. 

Successive dualities require the inherited interactions to remove the additional fields that would obstruct the next step. For $k=1$, the additional magnetic mesons carrying two gauge indices have $R$-charge $1$, which allows quadratic superpotential terms and can be integrated out at each $k=1$ step. Therefore, Seiberg duality can be applied successively to other gauge nodes. For $k\geq 2$, the proposed reduction uses
Higgsing and mass terms to eliminate the additional gauge–gauge mesons, while mixed gauge–flavor mesons can survive as ordinary quarks. The corresponding reductions are illustrated in
the two-node construction of ~\cite{Fang:2026ddf}.

\subsection{Intermediate and final dual gauge ranks}

\paragraph{Gauge ranks after one step}
At a selected node $i$, let $F_i^{\mathrm{eff}}$ be the number of ordinary flavor pairs entering the duality. This includes the original quarks and the surviving mixed gauge-flavor mesons inherited from the earlier steps. Such mesons are singlets under the previously dualized gauge group but act as quarks of the current node. The neighboring gauge groups supply a further $h_i=\sum_{j\sim i}N_j$ flavor pairs. For $k\ge2$, the KS duality therefore gives the magnetic gauge rank
\begin{equation}\label{eq:nodewise-local-rank}
 n_i=k\bigl(F_i^{\mathrm{eff}}+h_i\bigr)-N_i.
\end{equation}

Let $Y_i$ denote the magnetic adjoint, and let $u_i,\tilde u_i$ collect the magnetic bifundamentals connecting the dualized node to its neighbors before Higgsing. These are the magnetic quarks in the KS duality, rather than the electric fields $X_{ij},X_{ji}$.

We now explain how Higgsing is triggered.  Our superpotential contains a term of the form $\tilde u_iY_{i}u_{i}$, and it also incorporates the additional KS magnetic superpotential contribution built from electric singlets, along with the associated magnetic meson. The superpotentials that include the KS magnetic mesons are
\begin{equation}
    W\supset \sum_{s=0}^{k-1}[M_i^s] \tilde u_i Y_i^{k-1-s}u_i+Tr[M_i^1]-TrM_i^0\operatorname{diag}(X_j)_{j\sim i},\qquad 0\leq s\leq k-1.
\end{equation}
 
Here $[M_i^s]=[\tilde u_i X_i^s u_i]$ is the KS magnetic singlets and $\operatorname{diag}(X_j)_{j\sim i}$ is an $h_i\times h_i$ matrix. For simplicity, we refer to both electric and magnetic quarks simply as $u_i,\tilde u_i$. The F-term equations of KS mesons $[M_i^{s}]$ are
\begin{equation}\label{eq:gkr-moments}
 \widetilde u_iY_i^{s}u_i=-\delta_{s,k-2}\mathbf{1}_{h_i},\qquad 0\le s\le k-2;\qquad \tilde u_i Y_i^{k-1} u_i=\operatorname{diag}(X_j)_{j\sim i}.
\end{equation}
These equations require nonzero vacuum expectation values for $u_i$ and $\widetilde u_i$ when $k\geq 2$,
and for $Y_i$ when $k>2$. By contrast, other fields, such as $  X_j $, have vanishing vacuum expectation values $\langle X_j\rangle=0$ and are therefore neglected in the Higgsing discussion. Let $J_{k-1}$ be the $(k-1)\times(k-1)$ nilpotent Jordan matrix with ones on the superdiagonal, and let $e_a$ denote the standard basis column vectors. Writing $n_i=(k-1)h_i+m_i$, with $m_i\geq0$, a representative vacuum solution of the F-term equations is
\begin{equation}\label{eq:gkr-vacuum}
\begin{gathered}
 \langle Y_i\rangle=\begin{pmatrix}J_{k-1}\otimes\mathbf{1}_{h_i}&0\\0&0_{m_i}\end{pmatrix},\;\;
 \langle u_i\rangle=\begin{pmatrix}e_{k-1}\otimes\mathbf{1}_{h_i}\\0_{m_i\times h_i}\end{pmatrix},\;\;
\langle \tilde u_i\rangle=-\begin{pmatrix}e_1^T\otimes\mathbf{1}_{h_i}&0_{h_i\times m_i}\end{pmatrix}.
\end{gathered}
\end{equation}
The detailed Higgsing procedure is almost the same as those discussed in the two-node cases \cite{Fang:2026ddf}. Namely, after the Higgsing procedure and integrating out the massive chiral multiplets, those gauge-gauge mesons $[M_i^s]$ are eliminated, and the adjoint–bifundamental interactions \[
\operatorname{Tr}_i
\!\left(Y_i^{\rm new}X_{ij}^{\rm new}X_{ji}^{\rm new}\right)
\] are reproduced. Therefore, this Higgsing procedure applies at every step.

Hence, the remaining unbroken gauge groups are
\begin{equation}\label{eq:nodewise-higgs-rank}
 \begin{gathered}
 SU(n_i)\times\prod_{j\sim i}SU(N_j)
 \longrightarrow SU(m_i)\times\prod_{j\sim i}SU(N_j),\\
 \boxed{m_i=n_i-(k-1)h_i
 =kF_i^{\mathrm{eff}}+\sum_{j\sim i}N_j-N_i.}
 \end{gathered}
\end{equation}

For $k=1$, applying Seiberg duality yields the dual gauge rank
\begin{equation}\label{eq:nodewise-k1-rank}
 m_i=F_i^{\mathrm{eff}}+h_i-N_i,
\end{equation}
and no Higgsing is required. Both cases therefore lead to the same rank rule, with the appropriate
value of k. It remains to determine the ordinary flavors entering each step. The following example explains how inherited interactions affect this count.

\paragraph{An $A_2$ example: effective flavors $F^{\mathrm{eff}}$ in successive steps}
Consider $k=1$ and the three Seiberg steps $(1,2,1)$~\cite{Fang:2026ddf}. The original flavor families
$F_1,F_2$ retain their labels throughout. After the first step,
the magnetic $F_1$ pair $n_1,\widetilde n_1$ lies at node $1$,
and a new mixed $F_1$ pair $p_2,\widetilde p_2$ lies at node $2$.
Together with the original $F_2$ quarks, the latter supplies $F_2^{\mathrm{eff}}=F_1+F_2$ ordinary pairs for the second duality.

Write $X'_{12},X'_{21}$ for the bifundamentals after the first
step. The inherited cubic interactions include
\begin{equation}
 W\supset\operatorname{tr}(\widetilde n_1X'_{12}p_2)
  +\operatorname{tr}(\widetilde p_2X'_{21}n_1),
\end{equation}
with nonzero matching factors absorbed into the fields. When node
$2$ is dualized, $X'_{12}p_2$ and $\widetilde p_2X'_{21}$ become
the mixed mesons $m_1,\widetilde m_1$ at node $1$. The same terms then become quadratic terms:
\begin{equation}\label{eq:paperI-A2-flavor-mass}
 W\supset\operatorname{tr}(\widetilde n_1m_1+\widetilde m_1n_1).
\end{equation}
Both $F_1$ pairs at node $1$ are removed, and only $F_2$ is left. Thus, the actual flavor numbers entering the three
dualities are
\begin{equation}\label{eq:A2flavor-rank}
 F_1,\qquad F_1+F_2,\qquad F_2,
\end{equation}in order.
This example shows how mixed mesons supply new quarks while inherited interactions remove old ones. We next express these flavor changes in terms of root labels, providing a compact
description of the ADE rank transformations.

Before introducing the general flavor prescription, we recall the Weyl-group conventions used below.

\paragraph{Weyl-group definitions}
 
Let $\alpha_j$ be the simple root at node $j$, represented by the coordinate column $\boldsymbol e_j$ in the simple root basis, and let $\Phi_+$ be the set of positive roots. We write $s_i$ for an abstract simple reflection and $S_i$ for its matrix in this basis.
 
 The length of the Weyl group element \(w\in W\)is
\[
\boxed{
\ell(w)=\min\{m:\ w=s_{i_1}\cdots s_{i_m}\}.
}
\]
An expression containing exactly \(\ell(w)\) simple reflections is reduced, whereas an expression containing more than $\ell(w)$ simple reflections is nonreduced. Different expressions can represent the same element, whose length is always this minimum.

For example, in $ W(A_2)\), a reduced expression of $w$ is given by
\[
w=s_1s_2s_1=s_2s_1s_2,\qquad \ell(w)=3.
\]
We can insert arbitrarily many units $s_i^2=1$ to give nonreduced expressions:
\[
w=s_1s_2s_1(s_1s_1)^n.
\]
The displayed expression has \(3+2n\) factors, but the element still has length \(3\).

A finite Weyl group contains only finitely many elements, the length has a maximum. This maximum is attained by a unique element $w_0$, called longest element. 
\[
\boxed{
\ell(w_0)=\max_{w\in W}\ell(w)=|\Phi_+|.
}
\]
Moreover, the longest element maps each simple root to the negative of another simple root
\[
w_0(\Phi_+)=-\Phi_+, \qquad w_0(\alpha_j)=-\alpha_{\sigma(j)}.
\]
Here $\sigma$ is the opposition permutation, and its matrix $P$ is defined by $ P\boldsymbol e_j =\boldsymbol e_{\sigma(j)}.$ Thus the
matrix of $w_0$ in the simple root basis is $-P$. Since $w_0^2 =  e$, the opposition permutation satisfies $P^2 = I$. 

\paragraph{Root labels and effective flavors}
The effective flavors $F^{\mathrm{eff}}_i$ include the original $F_i$ quarks, as well as gauge–flavor mesons inherited from the previous steps, since any remaining mixed singlet is regarded as a quark of the current gauge node. To record
these labels, introduce root $\boldsymbol\beta_j$, written in simple root coordinates to each node, initially
$\boldsymbol\beta_j^{(0)}=\boldsymbol e_j$. Let
$G$ be the adjacency matrix and $C=2I_r-G$.  At step $t$, collect the current labels in \[
U_{t}
=\bigl(\boldsymbol\beta_1^{(t)},\ldots,
       \boldsymbol\beta_r^{(t)}\bigr),\qquad U_0=I_r,\qquad S_i=I_r-\boldsymbol e_i\boldsymbol e_i^TC.\]
If node $i$ is selected at this step, the transport rule is \begin{equation}\label{eq:recursive-formula}
\boldsymbol\beta_j^{(t)}
=
\boldsymbol\beta_j^{(t-1)}
-C_{ij}\boldsymbol\beta_i^{(t-1)}=(U_{t-1}S_iU_{t-1}^{-1})
\boldsymbol\beta_j^{(t-1)}.
\end{equation}

The node $i$ is fixed in this equation, rather than summing over $i$. Equivalently, 
\[
\boldsymbol\beta_j^{(t)}=
\begin{cases}
-\boldsymbol\beta_i^{(t-1)},&j=i,\\
\boldsymbol\beta_j^{(t-1)}+\boldsymbol\beta_i^{(t-1)},&j\sim i,\\
\boldsymbol\beta_j^{(t-1)},&\text{otherwise}.
\end{cases}
\]

This rule records the following changes in flavor representations:
\begin{itemize}
    \item At the dualized node \(j=i\):  magnetic quarks carry conjugate external flavor representations, recorded by $-\boldsymbol\beta_i^{(t-1)}$. This opposite sign convention also streamlines later mass cancellation: inherited cubic couplings turn into quadratic mass terms that pair appropriate old quarks with new mixed mesons in conjugate representations. Integrating out these massive pairs cancels oppositely signed flavor labels.
\item At a neighboring node \(j\sim i\): mixed gauge-flavor mesons supply new ordinary quarks with flavor label \(\boldsymbol\beta_i^{(t-1)}\), giving the result \(\boldsymbol\beta_j^{(t-1)}+\boldsymbol\beta_i^{(t-1)}\).
\item At all other nodes, the flavor label is unchanged.
\end{itemize}

With \(\boldsymbol\beta_j^{(0)}=\boldsymbol e_j\), the explicit formula can be derived from the recursive formula~\eqref{eq:recursive-formula}
\begin{equation}\label{eq:beta-generalt}
\boxed{
\boldsymbol\beta_j^{(t)}
=S_{i_1}S_{i_2}\cdots S_{i_t}\boldsymbol e_j
=U_t\boldsymbol e_j,
\qquad
U_t=S_{i_1}S_{i_2}\cdots S_{i_t},
}
\end{equation}

The flavor count entering step $t$ is determined by the label entering that step:
\begin{equation}\label{eq:nodewise-k1-actual-effective-flavors}
F_{i}^{t,\mathrm{eff}}
 =(\boldsymbol\beta_{i}^{(t-1)})^{T}\boldsymbol F.
\end{equation}

The subscript $i$ specifies the selected node.

The role of reducedness is now explicit. The Weyl-group length
criterion states that~\cite{Humphreys1990}
\[
 \ell(w s_i)=\ell(w)+1
 \quad\Longleftrightarrow\quad
 w(\alpha_i)\in\Phi_+.
\]
Every prefix of a reduced expression is reduced. Hence the root
selected before each step is positive. Its coefficients in the simple root basis are
nonnegative, so \eqref{eq:nodewise-k1-actual-effective-flavors}
gives a positive flavor count when all $F_i>0$. We use
reduced expressions below.

\paragraph{Root labels in the $A_2$ example.}
We can now compare the root prescription with the explicit flavor
count~\eqref{eq:A2flavor-rank}. For the sequence $(1,2,1)$,
\eqref{eq:recursive-formula} gives
Table~\ref{tab:nodewise-A2-roots}.
\begin{table}[H]
 \centering
 \renewcommand{\arraystretch}{1.15}
 \begin{tabular}{clcc}
 \toprule
Step $t$& State & $\boldsymbol\beta_1^{(t)}$ & $\boldsymbol\beta_2^{(t)}$\\
 \midrule
 0&Initially & \textcolor{red}{$\boldsymbol e_1$} & $\boldsymbol e_2$\\
1& After node $1$ & $-\boldsymbol e_1$ & \textcolor{red}{$\boldsymbol e_1+\boldsymbol e_2$}\\
2& After node $2$ & \textcolor{red}{$\boldsymbol e_2$ }& $-\boldsymbol e_1-\boldsymbol e_2$\\
3& After node $1$ & $-\boldsymbol e_2$ & $-\boldsymbol e_1$\\
 \bottomrule
 \end{tabular}
 \caption{ Root labels along the $A_2$
 sequence $(1,2,1)$, recorded after each step. Red entries are the labels selected for the next step.}
 \label{tab:nodewise-A2-roots}
\end{table}
For $A_2$, $C_{11}=C_{22}=2, C_{12}=C_{21}=-1$. First dualizing node \(i=1\), \((\boldsymbol \beta_1^{(0)},\boldsymbol \beta_2^{(0)})=(\boldsymbol e_1,\boldsymbol e_2)\), so $F_{1}^{1,\mathrm{eff}}=(\boldsymbol \beta_1^{(0)})^T\boldsymbol F=F_1$. After the first step, the updated labels are
\begin{equation}
\begin{aligned}
\boldsymbol \beta_1^{(1)}
&=\boldsymbol \beta_1^{(0)}-C_{11}\boldsymbol \beta_1^{(0)}
=\boldsymbol e_1-2\boldsymbol e_1=-\boldsymbol e_1,\\
\boldsymbol \beta_2^{(1)}
&=\boldsymbol \beta_2^{(0)}-C_{12}\boldsymbol \beta_1^{(0)}
=\boldsymbol e_2-(-1)\boldsymbol e_1
=\boldsymbol e_1+\boldsymbol e_2.
\end{aligned}
\end{equation}
 Next, we dualize node $i=2$, and use $ F_{2}^{2,\mathrm{eff}}=(\boldsymbol \beta_2^{(1)})^T\boldsymbol F=F_1+F_2$. After the second step,
\begin{equation}
\begin{aligned}
\boldsymbol \beta_1^{(2)}
&=\boldsymbol \beta_1^{(1)}-C_{12}\boldsymbol \beta_2^{(1)}
=-\boldsymbol e_1-(-1)(\boldsymbol e_1+\boldsymbol e_2)=\boldsymbol e_2,\\
\boldsymbol \beta_2^{(2)}
&=\boldsymbol \beta_2^{(1)}-C_{22}\boldsymbol \beta_2^{(1)}
=\boldsymbol e_1+\boldsymbol e_2-2(\boldsymbol e_1+\boldsymbol e_2)
=-(\boldsymbol e_1+\boldsymbol e_2).
\end{aligned}
\end{equation}
Finally, we dualize node $i=1$ again, with $ F_{1}^{3,\mathrm{eff}}=(\boldsymbol \beta_1^{(2)})^T\boldsymbol F=F_2$. After the third step, the final labels are
\begin{equation}
\begin{aligned}
\boldsymbol \beta_1^{(3)}
&=\boldsymbol \beta_1^{(2)}-C_{11}\boldsymbol \beta_1^{(2)}
=\boldsymbol e_2-2(\boldsymbol e_2)=-\boldsymbol e_2,\\
\boldsymbol \beta_2^{(3)}
&=\boldsymbol \beta_2^{(2)}-C_{12}\boldsymbol \beta_1^{(2)}
=-(\boldsymbol e_1+\boldsymbol e_2)-(-1)(\boldsymbol e_2)
=-\boldsymbol e_1.
\end{aligned}
\end{equation}
To summarize: at the beginning, the chosen labels are the positive simple roots $\boldsymbol e_1,\boldsymbol e_2$. At intermediate steps, the labels may be simple or non-simple, and they may be either positive or negative. In the end, the labels become negative simple roots, \(w_0(\alpha_j)=-\alpha_{\sigma(j)}\), which represent conjugate external flavor representations, not negative flavor multiplicities. The selected positive roots $\boldsymbol e_1$, $\boldsymbol e_1+\boldsymbol e_2$, and $\boldsymbol e_2$ reproduce the flavor counts $F_1,F_1+F_2,F_2$ obtained from the superpotential analysis.

\paragraph{Intermediate rank transformations}
Having determined the entering flavor labels, we return to the
gauge ranks. Neighboring gauge factors contribute their color ranks
as flavors, while only the selected gauge rank changes. The matrix
$S_i=I_r-\boldsymbol e_i\boldsymbol e_i^TC$ expresses the gauge contribution~\cite{CachazoEtAl2002}:
\[
 (S_i\boldsymbol N)_i=-N_i+\sum_{j\sim i}N_j,
 \qquad
 (S_i\boldsymbol N)_j=N_j\quad(j\neq i).
\]
Combining this action with
\eqref{eq:nodewise-k1-actual-effective-flavors}, the local rules
\eqref{eq:nodewise-higgs-rank} and \eqref{eq:nodewise-k1-rank}
take the common vector form
\begin{equation}\label{eq:nodewise-affine-rank}
 \boldsymbol N^{(a)}
 =S_{i_a}\boldsymbol N^{(a-1)}
 +k\boldsymbol e_{i_a}
  \bigl(\boldsymbol\beta_{i_a}^{(a-1)}\bigr)^T\boldsymbol F.
\end{equation}
The unit column $\boldsymbol e_{i_a}$ places the flavor contribution
only in the component corresponding to the dualized node.

We next express the rank after any number of steps directly in
terms of the initial data. Since
$U_a=U_{a-1}S_{i_a}$, $S_{i_a}^2=I_r$, and
$S_{i_a}\boldsymbol e_{i_a}=-\boldsymbol e_{i_a}$, we have
\[
 U_aS_{i_a}=U_{a-1},
 \qquad
 U_a\boldsymbol e_{i_a}
 =-\boldsymbol\beta_{i_a}^{(a-1)}.
\]
Multiplying \eqref{eq:nodewise-affine-rank} by $U_a$ therefore gives
\[
 U_a\boldsymbol N^{(a)}
 =U_{a-1}\boldsymbol N^{(a-1)}
 -k\boldsymbol\beta_{i_a}^{(a-1)}
   \bigl(\boldsymbol\beta_{i_a}^{(a-1)}\bigr)^T\boldsymbol F.
\]
Summing from $a=1$ to $t$ telescopes the rank terms. Using $U_0=I_r$
and multiplying by $U_t^{-1}$ yields
\begin{equation}\label{eq:nodewise-prefix-rank}
 \boxed{
 \boldsymbol N^{(t)}
 =U_t^{-1}\left[
 \boldsymbol N^{(0)}
 -k\sum_{a=1}^{t}
   \boldsymbol\beta_{i_a}^{(a-1)}
   \bigl(\boldsymbol\beta_{i_a}^{(a-1)}\bigr)^T\boldsymbol F
 \right].
 }
\end{equation}
The inverse is essential: a general intermediate product $U_t$
need not satisfy $U_t^2=I_r$.

\paragraph{Matching the simultaneous dual ranks}
 To reproduce the simultaneous rank map, the coefficient
of the initial ranks must become $-I_r$ after a final node
permutation. Comparing this requirement with $U_d^{-1}$ in
\eqref{eq:nodewise-prefix-rank} forces $U_d$ to be a negative
permutation matrix $U_d=-P$. Such a Weyl-group matrix sends every positive
root to a negative root and therefore represents $w_0$. Thus reproducing the full simultaneous rank map as an
identity in the initial ranks requires a word representing $w_0$.

Let $d$ denote the final step of such a sequence. Its final root
labels follow directly from \eqref{eq:beta-generalt}:
\[
 \boxed{
 \boldsymbol\beta_j^{(d)}
 =U_d\boldsymbol e_j
 =-P\boldsymbol e_j
 =-\boldsymbol e_{\sigma(j)}.
 }
\]
In particular, the endpoint labels are negative simple roots,
permuted by opposition. At this endpoint, unlike a general
intermediate step, $U_d^{-1}=U_d=-P$.

Now choose a \emph{reduced} expression of $w_0$. Its
$d=|\Phi_+|$ selected roots are positive, pairwise distinct, and
exhaust $\Phi_+$~\cite{Humphreys1990}. The ADE root identity gives
\[
 \sum_{a=1}^{d}
 \boldsymbol\beta_{i_a}^{(a-1)}
 \bigl(\boldsymbol\beta_{i_a}^{(a-1)}\bigr)^T
 =\sum_{\alpha\in\Phi_+}
   \boldsymbol\alpha\boldsymbol\alpha^T
 =hC^{-1}.
\]
Substituting this identity and $U_d^{-1}=-P$ into
\eqref{eq:nodewise-prefix-rank} gives
\begin{equation}\label{eq:nodewise-endpoint-rank}
 \boxed{
 \boldsymbol N_{\mathrm{end}}
 =P\bigl(khC^{-1}\boldsymbol F-\boldsymbol N^{(0)}\bigr).
 }
\end{equation}
After relabeling the endpoint nodes by opposition, the magnetic
ranks are therefore
\[
 \boldsymbol N^d
 =P\boldsymbol N_{\mathrm{end}}
 =khC^{-1}\boldsymbol F-\boldsymbol N^{(0)},
\]
in agreement with the simultaneous proposal~\eqref{eq:dual-ranks}.

\paragraph{Nodewise sequence as a finite-$N$ check}
No large-$N$ limit enters this nodewise derivation. The local rank rule
\eqref{eq:nodewise-affine-rank} and endpoint
\eqref{eq:nodewise-endpoint-rank} are exact integer identities. Also, the KS duality we used works for finite $N$. Therefore, this nodewise sequence realization can be viewed as a finite-$N$ consistency check of the ADE simultaneous duality proposal. Additional finite-$N$ checks are also presented in~\cite{Fang:2026ncq}.

\subsection{Minimal sequences: construction, equivalence, and counting}

In fact, there could be many nodewise sequences that achieve the simultaneous dualities. In this section, we aim to identify the nodewise sequence with minimal steps. At the Weyl-group level, each step corresponds to one simple reflection, and the full simultaneous transformation corresponds to \(w_0\).Consequently, a reduced expression of the Weyl element
\[
w_0=s_{i_1}\cdots s_{i_d}
\]
corresponds to the nodewise sequence that has the minimal steps.

\paragraph{A minimal sequence}

A minimal nodewise sequence realizing the longest Weyl-group
element has the form
\begin{equation}\label{eq:nodewise-sequence}
 \mathcal T_{\mathrm E}
 \xrightarrow{\mathrm{step}\ i_1}\mathcal T_1
 \xrightarrow{\mathrm{step}\ i_2}\cdots
 \xrightarrow{\mathrm{step}\ i_d}\mathcal T_d,
 \qquad d=|\Phi_+|=\frac{rh}{2},
\end{equation}
where $r$ is the number of gauge vertices, $h$ the Coxeter number.
Identifying $\mathcal T_d$ with the proposed magnetic theory establishes the simultaneous duality at finite
ranks. 

The mimimum number of steps is the standard length of $w_0$, as follows~\cite{Humphreys1990} 
\begin{equation}\label{eq:nodewise-step-count}
 d=\ell(w_0)=|\Phi_+|=\frac{rh}{2}
 =\begin{cases}
 r(r+1)/2,&\Gamma=A_r,\\
 r(r-1),&\Gamma=D_r,\\
 36,\ 63,\ 120,&\Gamma=E_6,\ E_7,\ E_8,\ \text{respectively}.
 \end{cases}
\end{equation}
This minimum counts steps represented by simple
reflections. To exhibit one such minimal ordering, let $ s_B$ and
$ s_W$ be the products of reflections at the black and white
vertices of the bipartite diagram.\footnote{A graph is bipartite
if its vertices admit two colors, with every edge joining vertices
of different colors. Finite ADE Dynkin diagrams are trees and hence
bipartite. Vertices of one color are nonadjacent, so their simple
reflections commute.} Reflections in each color class commute because they are not adjacent. The alternating construction gives
~\cite{BradyKennyWatt2010} 
\begin{equation}\label{eq:nodewise-bipartite-word}
 w_0=\begin{cases}
 (  s_B s_W)^{h/2},&h\ \text{even},\\
 ( s_B s_W)^{(h-1)/2} s_B,&h\ \text{odd}.
 \end{cases}
\end{equation}
These alternating expressions are reduced expressions of $w_0$: every selected root is positive before its step, and the final labels are negative.

For example, $A_3$ has $h=4$, with color classes $B=\{1,3\}$ and $W=\{2\}$, the corresponding expression is $w_0=(s_1s_3s_2)^2$,
which gives the node sequence $(1,3,2,1,3,2)$.

\paragraph{Equivalence of minimal sequences}
 
Having constructed one minimal sequence \eqref{eq:nodewise-bipartite-word}, we now
relate it to all other reduced expressions of $w_0$. For a simply laced diagram, the relevant
Coxeter relations are
\begin{equation}\label{eq:nodewise-coxeter-relations}
 \begin{aligned}
  s_i s_j&= s_j s_i,
 &&i\ne j,\quad i\not\sim j,\\
  s_i s_j s_i
 &= s_j s_i s_j,
 &&i\sim j.
 \end{aligned}
\end{equation}
The first exchanges nonadjacent reflections, the second is the
three-term braid relation at adjacent nodes. The word property for
Coxeter groups states that any two reduced expressions of the same
element are connected by these relations~\cite{Humphreys1990}.
Thus every minimal expression of $w_0$ can be obtained from the
chosen bipartite expression without changing its length or Weyl group endpoint.

For example, the $A_3$ sequence $(1,3,2,1,3,2)$ discussed above is related to the ordering $(1,2,3,1,2,1)$ by
\begin{equation}\label{eq:nodewise-A3-equivalence}
 \begin{aligned}
(1,3,2,1,3,2)&\longleftrightarrow(1,3,2,3,1,2) \\
 &\longleftrightarrow(1,2,3,2,1,2) \\
 &\longleftrightarrow(1,2,3,1,2,1).
 \end{aligned}
\end{equation}
Each line uses, respectively, the commutation of vertices $1$ and $3$,
the braid relation at $2,3$, and the braid relation at $1,2$.
All reduced expressions of $w_0$ therefore belong to one equivalence
class under \eqref{eq:nodewise-coxeter-relations}. They give the same
final root labels and the same endpoint rank
map~\eqref{eq:nodewise-endpoint-rank}, although their intermediate flavor multiplicities and gauge ranks can differ.

\paragraph{Counting minimal sequences}
We now count the distinct shortest ordered node sequences with the
same endpoint. Fix the node labels, and write $R(w)$ for the number
of reduced expressions of $w$. \footnote{Note that $R$ in this section is a counting function, not
an $R$-charge.} Equivalent words related by
\eqref{eq:nodewise-coxeter-relations} are counted separately when
their ordered node lists differ. The number is $R(w_0)$,
determined for every finite ADE type by \cite{Stanley1984}
\begin{equation}\label{eq:nodewise-word-recursion}
 R(e)=1,\qquad
 R(w)=\sum_{\{i:\ell(w s_i)=\ell(w)-1\}}R(w s_i).
\end{equation}
For example, in $A_2$,
\begin{equation}
 R(w_0)=R( s_1 s_2)
       +R( s_2 s_1)=1+1=2,
\end{equation}
corresponding to the two labeled words $(1,2,1)$ and $(2,1,2)$.

We now identify the shortest sequences related by a global
Dynkin-diagram automorphism, acting on every node label in the
word. Let $R_{\mathrm{sym}}(w_0)$ denote the resulting number of
symmetry classes.
\begin{equation}
 R_{\mathrm{sym}}(w_0)
 =\frac{R(w_0)}{|\operatorname{Aut}(\Gamma)|}.
\end{equation}
The automorphism group for $A_r$ ($r\ge2$),
$D_r$ ($r\ge5$), and $E_6$ is $\mathbb{Z}_2$ with order two. For $D_4$, the symmetric group $S_3$ has order six, and it is trivial for $A_1$, $E_7$, and $E_8$ with order one.

For instance, in type $A_r$ with $r\geq 2$, this result is provided in~\cite{Stanley1984}
\begin{equation}\label{eq:nodewise-word-count}
 R_{\mathrm{sym}}(w_0^{A_r})=
 \frac{\bigl[r(r+1)/2\bigr]!}
 {2\displaystyle\prod_{j=1}^{r}(2j-1)^{r+1-j}}.
\end{equation}

This quantity can become very large as we increase $r$. From a physical standpoint, such many different choices provide different nodewise duality sequences of the same simultaneous duality, as confirmed at the level of the ranks. Along the nodewise sequence, we also assume some consistency condition: the ranks of $F_i$ and $N_i$ should be selected so that the intermediate gauge rank remains positive and non-abelian. When the ranks of some or all of the dual gauge groups are one, it is also important to consider the corresponding s-confining cases. In that regime the dual pair can involve different numbers of gauge groups, which requires additional discussion that we leave for elsewhere.

\section{Subquiver composition}\label{sec:subquiver-composition}

In the previous section, we showed that simultaneous ADE dualities can be implemented as a sequence of dualities applied node by node. Here, we aim instead to represent these simultaneous ADE dualities as a sequence of simultaneous subquiver dualities. We focus on connected subquivers of an ADE quiver. Moreover, a simultaneous subquiver duality can itself be decomposed into nodewise dualities. In particular, the superpotential is fixed by the inherited nodewise superpotentials.

Similar to the nodewise sequence case, we can express the subquiver composition using the Weyl group. Let \(I\) denote the set of gauge vertices, and let \(r = |I|\) be the number of elements in \(I\).
For a connected proper vertex subset \(J\subsetneq I\), the standard parabolic subgroup
\[W_J=\langle s_j:j\in J\rangle\] is the Weyl group of the induced
subquiver $J$~\cite{Humphreys1990}. One simultaneous subquiver duality corresponds to the longest Weyl action element \(w_J\), which is called block. A reduced expression of this fixed
element uses the minimum number of nodewise steps
\[L_J=\ell(w_J)=|\Phi_J^+|=|J|h_J/2,\] where \(h_J\) is its Coxeter
number corresponding to the subquiver~\cite{Humphreys1990}.
Here \(\Phi_J^+\) is the set of positive roots of the subquiver. For the complete transformation and back to the simultaneous duality, we should require \(w=w_0\). For an ordered sequence of subquiver dualities, with blocks listed
in the order they are performed, the composition condition is
\begin{equation}\label{eq:subquiver-root-product}
 \boxed{w_0=w_{J_1}w_{J_2}\cdots w_{J_q}.}
\end{equation}
 
Here \(q\) denotes the number of blocks in the selected decomposition and
 can differ from one decomposition to another. In the nodewise case, each $J$ contains just a single node. For fixed $q$, only finitely many ordered block sequences are
possible, because the finite diagram has finitely many allowed
subsets $J$. If $q$ is not fixed, one can write infinitely many ordered factorizations, for instance by inserting units $s_i^2=1$:
\[
 w_0=w_{J_1}w_{J_2}\cdots w_{J_q}s_i^2.
\]
This does not change the Weyl group element, but it introduces two additional blocks. These are precisely the nonreduced expressions.

\paragraph{Nonreduced expansions and inverse steps.}
To distinguish these cases, suppose an expression of $w$ contains
$m=\ell(w)$ simple reflections and is therefore reduced. Appending
$s_i$ produces an expression with $m+1$ factors. This extended
expression is reduced precisely when
\[
 \ell(ws_i)=m+1=\ell(w)+1.
\]
The standard length criterion~\cite{Humphreys1990} states that
\[
 \ell(ws_i)=\ell(w)+1
 \quad\Longleftrightarrow\quad
 w(\alpha_i)\in\Phi_+,\qquad \alpha_i\text{ a simple root}.
\]
Thus a reduced expansion selects a positive root before every
step, giving $F_i^{t,\mathrm{eff}}>0$ under the flavor prescription
of the preceding section when all $F_j>0$.

A nonreduced expression can instead select a negative root. This
records a length-decreasing Weyl step and does not represent a negative physical flavor multiplicity. Let
$|\boldsymbol\beta_i|$ be the coordinate vector of the positive
root on the line spanned by $\boldsymbol\beta_i$. The proposed
extension to such steps is:
\begin{equation}\label{eq:nodewise-inverse-rank}
 F_i^{\mathrm{eff}}=|\boldsymbol\beta_i|^T\boldsymbol F,
 \qquad
 \boldsymbol N'=S_i\boldsymbol N
 +k\boldsymbol e_i|\boldsymbol\beta_i|^T\boldsymbol F.
\end{equation}
Its physical interpretation requires the step to realize an inverse
duality with the full inherited interactions included.

For an immediate second step at the same node,
$\boldsymbol\beta_i'=-\boldsymbol\beta_i$, so the entering count is
again $f=|\boldsymbol\beta_i|^T\boldsymbol F$. Since $S_i^2=I_r$
and $S_i\boldsymbol e_i=-\boldsymbol e_i$, the rank map is
involutive:
\begin{equation}\label{eq:nodewise-double-step}
 \boldsymbol N''
 =S_i(S_i\boldsymbol N+k\boldsymbol e_i f)
  +k\boldsymbol e_i f
 =\boldsymbol N.
\end{equation}
The matrix of root labels is recovered as well. These relations support the involutive nature of Seiberg dualities at the level of roots and ranks.

For a nonreduced expansion, this effective flavor \eqref{eq:nodewise-inverse-rank} modifies the dual rank formula \eqref{eq:nodewise-prefix-rank} to
\begin{equation}\label{eq:subquiver-signed-rank-sum}
 U_t\boldsymbol N^{(t)}
 =\boldsymbol N^{(0)}
 -k\sum_{a=1}^{t}
 \boldsymbol\beta_{i_a}^{(a-1)}
 \bigl|\boldsymbol\beta_{i_a}^{(a-1)}\bigr|^T\boldsymbol F.
\end{equation}
Because $|\beta_i|$ is a positive root and $\sum_{\alpha\in\Phi_+}\boldsymbol\alpha\boldsymbol\alpha^T = hC^{-1}$, this extension yields the same formal endpoint ranks~\eqref{eq:nodewise-endpoint-rank}, including in the nonreduced cases presented below. 

We first illustrate a reduced regrouping in $A_3$, retaining the
intermediate spectrum. We then present factorizations in the
remaining ADE families whose elementary expansions are generally
nonreduced.

\paragraph{$A_3$ example}
Set \(k=1\) for the \(A_3\) gauge quiver. Each step is Seiberg duality followed by integrating out massive fields. The reduced word \(123121\) can be partitioned as \( (1)(2)(3)(121)\). The corresponding Weyl group element admits the reduced decomposition
\begin{equation}\label{eq:subquiver-A3-grouping}
 w_0(A_3)=s_1s_2s_3w_{\{1,2\}},\qquad
 w_{\{1,2\}}=s_1s_2s_1.
\end{equation}
The first three blocks are nodewise dualities on nodes \(1,2,3\), respectively. The final \(A_2\) block consists of the three steps \(121\). We can check that the magnetic theory obtained after four blocks is consistent with the simultaneous magnetic theory shown in Figure~\ref{fig:paperI-A3-subquiver-spectrum}.
\begingroup
\captionsetup[figure]{font=footnotesize,skip=1mm}
\captionsetup[subfigure]{font=scriptsize,labelfont=bf,
 justification=centering,singlelinecheck=false,skip=.4mm,hypcap=true}
\newcommand{\PaperIAThreeGraph}[5]
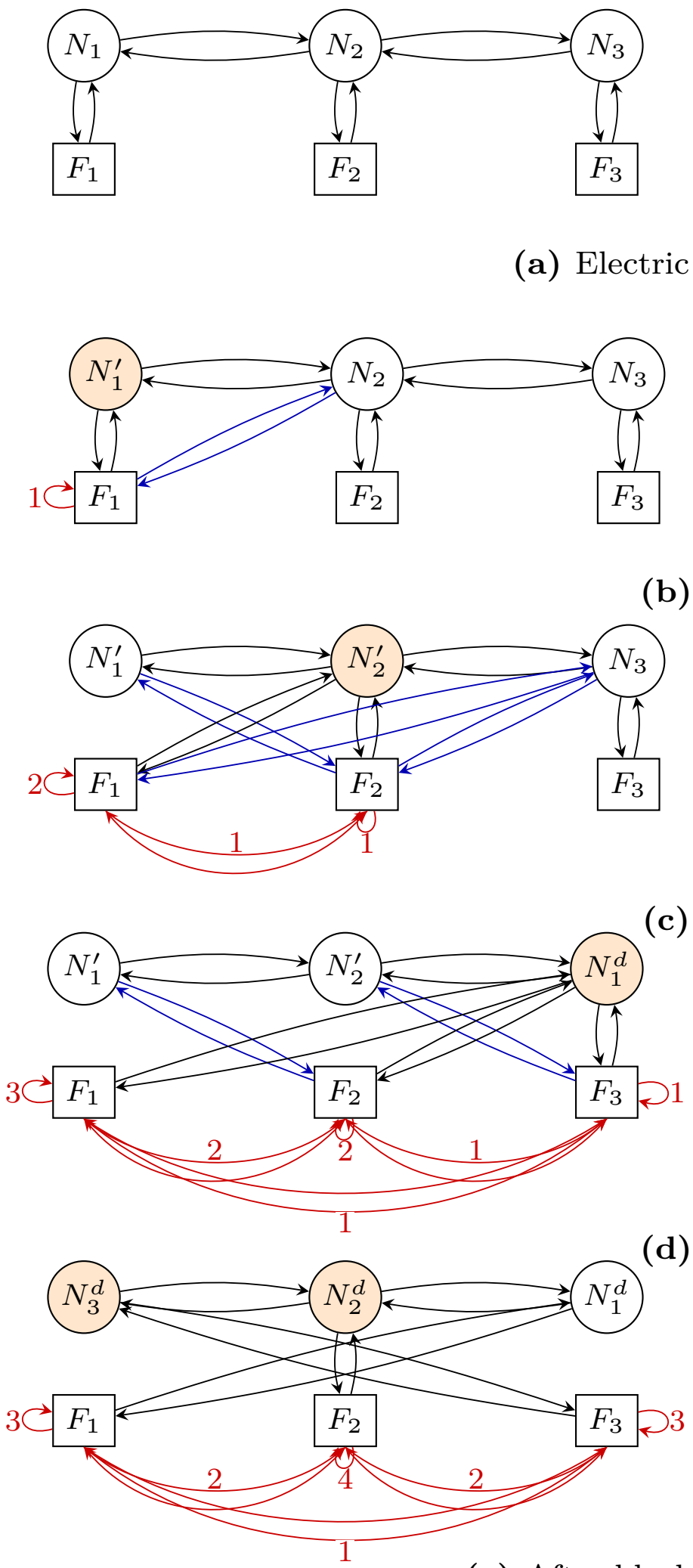
{%
\begin{tikzpicture}[x=.88mm,y=.72mm,
 every node/.style={font=\fontsize{7.5}{8}\selectfont},
 gauge/.style={circle,draw,minimum size=6.3mm,inner sep=.4pt},
 flavor/.style={rectangle,draw,fill=white,minimum width=5.4mm,
   minimum height=4.5mm,inner sep=.4pt},
 chiral/.style={->,>=stealth,line width=.35pt},
 singlet/.style={chiral,red!80!black}]
 \foreach \i/\xx/\rk in {1/0/#2,2/26/#3,3/52/#4}{%
  \def\AThreeFill{white}%
  \ifnum#1>0
   \ifnum#1<4
    \ifnum#1=\i \def\AThreeFill{orange!20}\fi
   \else
    \ifnum\i<3 \def\AThreeFill{orange!20}
    \fi
   \fi
  \fi
  \node[gauge,fill=\AThreeFill] (g\i) at (\xx,0) {$\rk$};
  \node[flavor] (f\i) at (\xx,-15) {$F_{\i}$};
 }
 \foreach \i/\j in {1/2,2/3}{%
  \draw[chiral] (g\i) to[bend left=9] (g\j);
  \draw[chiral] (g\j) to[bend left=9] (g\i);
 }
 #5
\end{tikzpicture}%
}
\newcommand{\PaperIAThreePair}[3]{%
 \ifnum#1=#2
  \draw[chiral,draw=#3] (g#1) to[bend right=12] (f#2);
  \draw[chiral,draw=#3] (f#2) to[bend right=12] (g#1);
 \else
  \draw[chiral,draw=#3] (g#1) to[bend left=5] (f#2);
  \draw[chiral,draw=#3] (f#2) to[bend left=5] (g#1);
 \fi
}
\newcommand{\PaperIAThreeLoop}[2]{%
 \ifnum#1=1
  \path[singlet] (f#1) edge[loop left,looseness=6]
    node[left,inner sep=.4pt] {$#2$} (f#1);
 \else\ifnum#1=2
  \path[singlet] (f#1) edge[loop below,looseness=5]
    node[below,inner sep=.4pt] {$#2$} (f#1);
 \else
  \path[singlet] (f#1) edge[loop right,looseness=6]
    node[right,inner sep=.4pt] {$#2$} (f#1);
 \fi\fi
}
\newcommand{\PaperIAThreeSinglets}[3]{%
 \ifnum#1=1\ifnum#2=3
  \draw[singlet] (f#1.south) .. controls +(15,-15) and +(-15,-15) ..
   node[below,pos=.5,fill=white,inner sep=.3pt] {$#3$} (f#2.south);
  \draw[singlet] (f#2.south) .. controls +(-15,-12) and +(15,-12) .. (f#1.south);
 \else
  \draw[singlet] (f#1.south) .. controls +(7,-7) and +(-7,-7) ..
   node[above,pos=.5,fill=white,inner sep=.3pt] {$#3$} (f#2.south);
  \draw[singlet] (f#2.south) .. controls +(-7,-10) and +(7,-10) .. (f#1.south);
 \fi\else
  \draw[singlet] (f#1.south) .. controls +(7,-7) and +(-7,-7) ..
   node[above,pos=.5,fill=white,inner sep=.3pt] {$#3$} (f#2.south);
  \draw[singlet] (f#2.south) .. controls +(-7,-10) and +(7,-10) .. (f#1.south);
 \fi
}
\begin{figure}[H]
\centering
\begin{subfigure}{\linewidth}
\centering
\begin{minipage}[c][24mm][c]{.43\linewidth}
\centering
\PaperIAThreeGraph{0}{N_1}{N_2}{N_3}{%
 \PaperIAThreePair{1}{1}{black}
 \PaperIAThreePair{2}{2}{black}
 \PaperIAThreePair{3}{3}{black}
}
\end{minipage}\hfill
\begin{minipage}[c]{.54\linewidth}
\centering
{\fontsize{8.5}{10}\selectfont\setlength{\arraycolsep}{2.4pt}
$\mathscr S^{(0)}(y)=0_{3\times3}$
\par\vspace{.7mm}New singlets: 0; Total singlets: 0.
}
\end{minipage}
\caption{Electric theory: no elementary singlets.}
\label{fig:paperI-A3-block-stage-0}
\end{subfigure}\par\vspace{.3mm}

\begin{subfigure}{\linewidth}
\centering
\begin{minipage}[c][24mm][c]{.43\linewidth}
\centering
\PaperIAThreeGraph{1}{N_1'}{N_2}{N_3}{%
 \PaperIAThreePair{1}{1}{black}
 \PaperIAThreePair{2}{1}{blue!70!black}
 \PaperIAThreePair{2}{2}{black}
 \PaperIAThreePair{3}{3}{black}
 \PaperIAThreeLoop{1}{1}
}
\end{minipage}\hfill
\begin{minipage}[c]{.54\linewidth}
\centering
{\fontsize{8.5}{10}\selectfont\setlength{\arraycolsep}{2.4pt}
$\mathscr S^{(1)}(y)=\begin{pmatrix}
1&0&0\\
0&0&0\\
0&0&0
\end{pmatrix}$
\par\vspace{.7mm}New singlets: 1; Total singlets: 1.
}
\end{minipage}
\caption{After block $1$: node $1$.}
\label{fig:paperI-A3-block-stage-1}
\end{subfigure}\par\vspace{.3mm}

\begin{subfigure}{\linewidth}
\centering
\begin{minipage}[c][24mm][c]{.43\linewidth}
\centering
\PaperIAThreeGraph{2}{N_1'}{N_2'}{N_3}{%
 \PaperIAThreePair{1}{2}{blue!70!black}
 \PaperIAThreePair{2}{1}{black}
 \PaperIAThreePair{2}{2}{black}
 \PaperIAThreePair{3}{1}{blue!70!black}
 \PaperIAThreePair{3}{2}{blue!70!black}
 \PaperIAThreePair{3}{3}{black}
 \PaperIAThreeLoop{1}{2}
 \PaperIAThreeLoop{2}{1}
 \PaperIAThreeSinglets{1}{2}{1}
}
\end{minipage}\hfill
\begin{minipage}[c]{.54\linewidth}
\centering
{\fontsize{8.5}{10}\selectfont\setlength{\arraycolsep}{2.4pt}
$\mathscr S^{(2)}(y)=\begin{pmatrix}
1+y^{2}&y&0\\
y&1&0\\
0&0&0
\end{pmatrix}$
\par\vspace{.7mm}New singlets: 4; Total singlets: 5.
}
\end{minipage}
\caption{After block $2$: node $2$.}
\label{fig:paperI-A3-block-stage-2}
\end{subfigure}\par\vspace{.3mm}

\begin{subfigure}{\linewidth}
\centering
\begin{minipage}[c][24mm][c]{.43\linewidth}
\centering
\PaperIAThreeGraph{3}{N_1'}{N_2'}{N_1^d}{%
 \PaperIAThreePair{1}{2}{blue!70!black}
 \PaperIAThreePair{2}{3}{blue!70!black}
 \PaperIAThreePair{3}{1}{black}
 \PaperIAThreePair{3}{2}{black}
 \PaperIAThreePair{3}{3}{black}
 \PaperIAThreeLoop{1}{3}
 \PaperIAThreeLoop{2}{2}
 \PaperIAThreeLoop{3}{1}
 \PaperIAThreeSinglets{1}{2}{2}
 \PaperIAThreeSinglets{2}{3}{1}
 \PaperIAThreeSinglets{1}{3}{1}
}
\end{minipage}\hfill
\begin{minipage}[c]{.54\linewidth}
\centering
{\fontsize{8.5}{10}\selectfont\setlength{\arraycolsep}{2.4pt}
$\mathscr S^{(3)}(y)=\begin{pmatrix}
1+y^{2}+y^{4}&y+y^{3}&y^{2}\\
y+y^{3}&1+y^{2}&y\\
y^{2}&y&1
\end{pmatrix}$
\par\vspace{.7mm}New singlets: 9; Total singlets: 14.
}
\end{minipage}
\caption{After block $3$: node $3$.}
\label{fig:paperI-A3-block-stage-3}
\end{subfigure}\par\vspace{.3mm}

\begin{subfigure}{\linewidth}
\centering
\begin{minipage}[c][24mm][c]{.43\linewidth}
\centering
\PaperIAThreeGraph{6}{N_3^d}{N_2^d}{N_1^d}{%
 \PaperIAThreePair{1}{3}{black}
 \PaperIAThreePair{2}{2}{black}
 \PaperIAThreePair{3}{1}{black}
 \PaperIAThreeLoop{1}{3}
 \PaperIAThreeLoop{2}{4}
 \PaperIAThreeLoop{3}{3}
 \PaperIAThreeSinglets{1}{2}{2}
 \PaperIAThreeSinglets{2}{3}{2}
 \PaperIAThreeSinglets{1}{3}{1}
}
\end{minipage}\hfill
\begin{minipage}[c]{.54\linewidth}
\centering
{\fontsize{8.5}{10}\selectfont\setlength{\arraycolsep}{2.4pt}
$\mathscr S^{(6)}(y)=\begin{pmatrix}
1+y^{2}+y^{4}&y+y^{3}&y^{2}\\
y+y^{3}&1+2y^{2}+y^{4}&y+y^{3}\\
y^{2}&y+y^{3}&1+y^{2}+y^{4}
\end{pmatrix}=\mathcal P_8^{A_3}(y)$
\par\vspace{.7mm}New singlets: 6; Total singlets: 20.
}
\end{minipage}
\caption{After block $4$:on node $1$ and $2$ simultaneously.}
\label{fig:paperI-A3-block-stage-6}
\end{subfigure}\par\vspace{.3mm}

\caption{The $k=1$ block sequence $(1)(2)(3)(121)$. Orange marks the blocks that were dualized in the previous step. Black and blue arrow pairs denote quarks and mixed mesons, red arrows and loops denote singlets. A coefficient $c y^d$ in $\mathscr S^{(t)}_{ij}$ gives $c$ species of charge $R(Q_i)+R(Q_j)+d/2$, each with $F_iF_j$ components, with $i,j=1,2,3$. Each step creates singlets from all ordered pairs of entering
ordinary flavors~\cite{Seiberg:1994pq}. The numbers for their species are $1,4,9,6$. The endpoint matrix is $\mathcal P_8^{A_3}(y)$, with twenty species. The intermediate dual ranks are $N_1'=F_1+N_2-N_1, N_2'=2F_1+F_2+N_3-N_1, N_1''=F_1+2F_2+N_3-N_2.$ The endpoint ranks are $(N_3^d,N_2^d,N_1^d)$, where $\boldsymbol N^d=4C_{A_3}^{-1}\boldsymbol F-\boldsymbol N$.}
\label{fig:paperI-A3-subquiver-spectrum}
\end{figure}
\endgroup

Next, we provide further examples of subquiver dualities without repeating the detailed analysis carried out for the $A_3$ example. 

\paragraph{Type \(A_r\).}
Label the chain \(1-\cdots-(r)\). For \(A_r\), \(r\ge3\), let \([a,b]\)
denote the gauge node interval from \(a\) to \(b\). The explicit
four-block decomposition and the subquiver duality sequence are
\begin{equation}\label{eq:subquiver-four-A}
 \begin{aligned}
 w_0(A_r)&=w_{[3,r]}w_{[2,r-1]}w_{[1,r-2]}w_{[r-1,r]},\quad\hspace{1em}(A_{r-2},A_{r-2},A_{r-2},A_2).
 \end{aligned}
\end{equation}
The entries in parentheses provide the successive subquiver types. For example, in the case of \(A_3\), these are \((A_1,A_1,A_1,A_2)\), which point back to the previous $A_3$ example.

\paragraph{Type \(D_r\)}
Label the chain \(1-\cdots-(r-1)\) with the extra edge
\((r-2,r)\), and let \(I=\{1,\ldots,r\}\) be the full set of
gauge vertices. The three-block decomposition and the corresponding subquiver duality sequence are
\begin{equation}\label{eq:subquiver-D-maximal-blocks}
 \begin{gathered}
 w_0(D_r)=w_{I\setminus\{r\}}w_{I\setminus\{r-1\}}w_{I\setminus\{1\}}\qquad(A_{r-1},A_{r-1},D_{r-1}).
 \end{gathered}
\end{equation}
For example, at \(r=4\),
\(D_3=A_3\), the three block types are \((A_3,A_3,A_3)\).

\paragraph{Exceptional types $E_r,\,r=6,7,8$}
Use the chain \(1-\cdots-(r-1)\) with the extra edge \((3,r)\).
Put \(B_r=w_{I\setminus\{r-1\}}\),
\(F_r=w_{I\setminus\{1\}}\), and
\(T_8=w_{\{2,3,4,5,6,8\}}\). Representatives are
\begin{equation}\label{eq:subquiver-E-minimal-words}
 \begin{aligned}
&w_0( E_6)= B_6F_6B_6 &&(D_5,D_5,D_5),\\
&w_0(  E_7)=s_6B_7F_7B_7 &&(A_1,E_6,D_6,E_6),\\
&w_0(  E_8)=T_8B_8F_8B_8F_8 &&(D_6,E_7,D_7,E_7,D_7).
 \end{aligned}
\end{equation}
To summarize, the dualities associated with a large $A_r$-type quiver can be reduced to dualities on smaller $A_m$ subquivers with $m\leq r$. Similarly, $D_r$ dualities can be split into a $D_{r-1}$ sector plus additional $A$-type dualities; $E_6$ may be assembled from $D$- and $A$-type building blocks; $E_7$ from $A$, $D$, and $E_6$ components; and $E_8$ from $E_7$ together with suitable $D$-type contributions. The relevant decompositions are summarized in the following table:
\begin{center}
\begin{tabular}{lcc}
\toprule
Diagram &\(\sum_a\ell(w_{J_a})\) & \(\ell(w_0)\)\\
\midrule
\(A_r\), \(r\ge4\)
& \(\frac{3(r-2)(r-1)}{2}+3\)
& \(\frac{r(r+1)}{2}\)\\
\(D_r\), \(r\ge4\)
& \(2(r-1)^2\)
& \(r(r-1)\)\\
\(E_6\) & \(60\)  & \(36\)\\
\(E_7\) & \(103\) & \(63\)\\
\(E_8\) & \(240\) & \(120\)\\
\bottomrule
\end{tabular}
\end{center}  
For each case in this table,
\[
 \sum_{a=1}^{q}\ell(w_{J_a})>\ell(w_0),
\]
 Consequently, every elementary expansion listed in the table is nonreduced, even though every individual block has been expanded in a reduced expression.

\section{Conclusion}
\label{sec:conclusion}
In this paper, we use the large-$N$ index approach to identify a set of new simultaneous quiver dualities that admit a finite meson basis. For a particular choice of $R$-charge assignment, the resulting graphs correspond to finite ADE diagrams.

 The strategy asks whether there exists an $L>0$ for which
$\mathcal P_L$ is a finite polynomial with nonnegative integer
coefficients. It does not uniquely determine \(L\) since there is an identity 
\begin{equation}\label{eq:meson-polynomial-multiples}
 \mathcal P_{mL}(y)
 =\left(\sum_{a=0}^{m-1}y^{aL}\right)\mathcal P_L(y),
 \qquad m\in\mathbb Z_{>0}.
\end{equation}
Thus every positive integer multiple of an admissible $L$ is
also admissible. Since $\mathcal P_{mL}(1)=m\mathcal P_L(1)$,
these choices have different meson multiplicities.

For the specific ADE superpotential chosen in
Section~\ref{sec:adeproposal}, Appendix~\ref{app:meson-graded-matching} derives the classical meson Hilbert series $\operatorname{Hilb}(\mathcal{B}_k;y)$ from the quotient path algebra $\mathcal B_k$ defined by the F-term equations. Matching this meson spectrum with the index prediction \[\text{Hilb}(\mathcal{B}_k;y)=\mathcal{P}_{2hk}(y)\]
selects $L=2kh$ for the proposal studied here. Larger multiples of $2hk$
still satisfy the large-$N$ index strategy but predict additional independent meson species that
are absent from the classical spectrum of this superpotential. 

As a finite-$N$ check, we show that these simultaneous ADE dualities can be realized by a sequence of known nodewise dualities. We organized the proposed ADE nodewise duality sequences using reduced expressions of the longest Weyl group element. Under the
conditions stated in Section~\ref{sec:nodewise-sequence}, these
sequences reproduce the simultaneous rank map. These simultaneous ADE dualities may likewise be understood as a sequence of simultaneous dualities of connected subquivers, obtained from reduced or nonreduced expressions of the corresponding longest Weyl group element. This paper emphasizes the Weyl group organization of the rank transformations and does not provide a detailed discussions of superpotentials. More extensive discussions of the two-node duality appear in the companion work~\cite{Fang:2026ddf}.

A few questions remain open, as outlined below.

In this work, we focus on two particular choices of $R$-charges for quivers. These correspond to special simultaneous dualities exhibiting a classical truncation of the chiral ring. More general quivers with arbitrary $R$-charge assignments can still be studied using the large-$N$ index strategy derived in \cite{Fang:2026ncq}, where quantum chiral ring constraints may need to be added in general. The ADE simultaneous dualities developed here admit nodewise realizations, but it remains an open question whether more general classes of simultaneous quiver dualities can be realized in the same way.

Besides the quivers that include an adjoint chiral multiplet at each gauge node, or those with no adjoint fields, there are situations in which partial adjoints are absent. Then the graph allows for more configurations, depending on the choice of $k$.  Also, throughout this paper, every gauge node carries \(F_i>0\) flavors. Allowing flavor groups at only a subset of gauge nodes provides a further extension. It is also natural to explore extensions involving orthogonal or symplectic gauge groups, as well as three-dimensional counterparts.

Additional dynamical problems remain to be studied. For instance, one-loop asymptotically free regions (\ref{eq:dyn-af}) do not necessarily guarantee an interacting infrared
fixed point: couplings to the other gauge factors and to the superpotential can instead drive a gauge coupling to zero in the infrared, even if it is asymptotically free at one loop. With two
gauge factors, there are four formal possibilities for the zero/nonzero fixed-point couplings, depending
on whether each coupling vanishes or stays nonzero~\cite{Barnes:2005zn}. More generally, an ADE quiver with $r$ gauge factors
 admits $2^r$ such formal patterns. 
A further direction is to study RG flows and superpotential relevance
using $a$-maximization, following analyses of theories with a single
gauge group~\cite{Kutasov:2003iy,Intriligator:2003mi,Cho:2024Landscape}.
Such an analysis should also take into consideration accidental symmetries arising from gauge-invariant chiral operators that decouple and become free fields~\cite{Kutasov:2003iy,BenvenutiGiacomelli2017ChiralRing}.

In contrast to the field theoretic approach, brane constructions could provide a geometric interpretation of the nodewise sequences. Brodie and Hanany constructed product group brane dualities for general $k$ and an explicit nodewise sequence at $k=1$~\cite{Brodie:1997sz}. Since Seiberg duality, Kutasov--Schwimmer duality, and Higgsing each admit brane realizations~\cite{arXiv:hep-th/9802067,arXiv:hep-th/9702014,arXiv:hep-th/9704104}, one can ask whether successive configurations reproduce the proposed general-$k$ transformations and how this construction extends to general ADE quivers.

From a mathematical standpoint, Berenstein and Douglas connected quiver Seiberg duality with tilting equivalences between derived categories of quiver algebras~\cite{arXiv:hep-th/0207027}. It would be interesting to investigate whether the proposed simultaneous ADE dualities, together with their nodewise sequences, admit a similar categorical description and reveal hidden mathematical links between them.
\acknowledgments
We thank Dan Xie, Sungjoon Kim, and Tadashi Okazaki for inspiring discussions and related collaborations. We also note that AI tools were used to assist in drafting and polishing the mathematical proofs in the appendices. The author has independently checked and confirmed all results. This work is supported in part by the National Natural Science Foundation of China(NSFC) through the Research Fund for International Excellent Young Scientists(Grant No. W2632007).

 \appendix

\section{Classification of quivers without adjoint matter}
\label{app:non-adj-proof}
This appendix proves the graph classifications for
quivers without adjoint under the finite meson requirement when every gauge node has ordinary flavors, which is summarized in Table \ref{tab:non-adj-class}.

The $k=1$ case coincides with the
case of a quiver with an adjoint chiral multiplet at every gauge node, as proven in section~\ref{sec:full-adjoint-criterion}.

For $k\geq2$, polynomiality imposes a strong restriction
on the size and connectivity of the gauge graph. Let
$\lambda_{\max}$ be the largest eigenvalue of $G_\Gamma$.
On a corresponding eigenvector, the matrix in brackets in
\eqref{eq:no-adjoint-index-kernel} acts as the scalar polynomial
\begin{equation}
 d(y)=\sum_{a=0}^{k}y^{2a}
       -\lambda_{\max}y\sum_{a=0}^{k-1}y^{2a},
 \qquad d(0)=1,\qquad d(1)=k+1-k\lambda_{\max}.
\end{equation}
The corresponding eigenvalue of $\mathcal P_L(y)$ is
\begin{equation}\frac{(1-y^L)}{(1-y^2)d(y)}.\end{equation} If $d(1)<0$, continuity forces
$d(y)$ to vanish at some $0<y<1$, where $1-y^L$ is
nonzero. If $d(1)=0$, the denominator has a zero of order
at least two at $y=1$, whereas the numerator has only a
simple zero. Both cases leave an uncanceled pole.
Polynomiality therefore requires $d(1)>0$, equivalently,
\begin{equation}
\lambda_{\max}(G_\Gamma)<1+\frac1k.
\end{equation}

By Proposition~3.1.1 of Brouwer--Haemers~\cite{BrouwerHaemers2012},
removing vertices or edges cannot increase the largest
adjacency eigenvalue. Thus, for every subgraph $T\subseteq\Gamma$,
\begin{equation}
 \lambda_{\max}(G_T)\leq\lambda_{\max}(G_\Gamma).
\end{equation}
Therefore, we begin with a small number of vertices and show that cases with more vertices are not allowed.

For two vertices, the only connected simple graph is
$A_2$. Its largest adjacency eigenvalue is $1$, so it
passes the necessary bound for every $k\geq2$.
Two vertices joined by $m\geq2$ edges have adjacency matrix
\begin{equation}
\begin{pmatrix}0&m\\m&0\end{pmatrix},
\end{equation}
whose eigenvalues are $m$ and $-m$. Any graph containing this pair must satisfy
$\lambda_{\max}(G_\Gamma)\geq m\geq2$ and violate the bound. Therefore, only one edge is permitted. As a result, in a graph with more vertices, there is still at most one edge between any two vertices.

For three vertices, a connected simple graph is either
the path $A_3$ or a triangle. The triangle has largest
adjacency eigenvalue two and is excluded. Since
\begin{equation}
 \lambda_{\max}(A_3)=\sqrt2<\frac32,
 \qquad \sqrt2>\frac43,
\end{equation}
$A_3$ passes the necessary bound at $k=2$ and fails
for every $k\geq3$.

For four or more vertices, connectedness guarantees a
four-vertex tree subgraph: start at any vertex and attach
an adjacent new vertex at each step until four vertices
have been reached, retaining only the three chosen edges.
The resulting tree $T$ is either the path $A_4$, if every
vertex has degree at most two, or the star graph
$K_{1,3}$, if one vertex has degree three. Any extra
edges may be discarded since the subgraph need not
be induced. The two possibilities have
\begin{equation}
 \lambda_{\max}(A_4)=\frac{1+\sqrt5}{2}>\frac32,
 \qquad
 \lambda_{\max}(K_{1,3})=\sqrt3>\frac32.
\end{equation}
The subgraph comparison therefore gives
$\lambda_{\max}(G_\Gamma)>3/2$, excluding every connected
graph with four or more vertices for all $k\geq2$.
Thus the only remaining candidates are $A_2$ for every
$k\geq2$ and $A_3$ at $k=2$.

To prove sufficiency, put $S_j=\sum_{a=0}^{j}y^{2a}$.
Direct inversion gives
\begin{equation}
 \mathcal P_{4k+2}^{A_2}(y)
 =\begin{pmatrix}
 S_k&yS_{k-1}\\
 yS_{k-1}&S_k
 \end{pmatrix},
\end{equation}
and, for $A_3$ at $k=2$,
\begin{equation}
 \mathcal P_{24}^{A_3}(y)
 =(1+y^4)(1+y^6)
 \begin{pmatrix}
 S_4&y(1+y^2)S_2&y^2(1+y^2)^2\\
 y(1+y^2)S_2&S_2^2&y(1+y^2)S_2\\
 y^2(1+y^2)^2&y(1+y^2)S_2&S_4
 \end{pmatrix}.
\end{equation}
Every coefficient is a nonnegative integer, and direct
multiplication verifies
\begin{equation}
\mathcal P_L(y)=(1-y^L) H(Q,y^{k+1}).
\end{equation}
Together with the $k=1$ result, these polynomials complete
the classification in Table~\ref{tab:non-adj-class}.

\section{Meson operators and graded counting}\label{sec:appendixofalgebra}

In this Appendix,  we first determine a complete
independent basis \eqref{eq:meson-spectrum}, then compute the corresponding Hilbert series, and finally prove that it agrees
with the matrix $\mathcal P_{2kh}(y)$ predicted by the large-$N$ index strategy. 

\paragraph{Path algebra modulo F-term equations}
 Consider the gauge quiver part of the electric theory. Let $\overline\Gamma$ be the doubled ADE quiver, containing a pair of bifundamental arrows on every edge, and let $\Gamma_{adj}$ also include an adjoint loop $X_i$ at each vertex. The path algebra consists of linear combinations of paths, with multiplication given by joining paths at a common endpoint. If the endpoints do not match, their product is zero. The length-zero paths $e_i$ are vertex projectors satisfying $e_i e_j=\delta_{ij}e_i$ and $\sum_i e_i=1$. In particular, $X_i=e_iX_ie_i$ and $X_{ij}=e_iX_{ij}e_j$.
  
The F-term relations  define the quotient path algebra \(\mathcal B_k\)
\begin{equation}
\mathcal B_k=
\frac{\mathbb C \Gamma_{\mathrm{adj}}}
{\left\langle
X_iX_{ij}-X_{ij}X_j,\quad
\mu_i+X_i^k
\right\rangle}.
\end{equation}
Here $\mu_i=\sum_{(i,j)\in\mathcal O}X_{ij}X_{ji}
       -\sum_{(j,i)\in\mathcal O}X_{ij}X_{ji}$ is defined in section \ref{sec:adeproposal}, and the brackets denote the ideal generated by the displayed F-term relations.

With this algebra, the adjoints combine into the central element, and it commutes with other paths \[Z=\sum_i X_i,\qquad Ze_i=e_iZ=X_i,\qquad ZX_{ij}=X_i X_{ij}=X_{ij}X_j=X_{ij}Z.\]

Multiplication by the projectors selects the endpoints, so $$V_k^{ij}=e_i\mathcal B_ke_j$$ is the vector space of path classes from $j$ to $i$.
The weighted grading is
\begin{equation}
\deg X_{ij}=k,
\qquad
\deg X_i=\deg Z=2.
\end{equation}
Both terms in F-term relations are homogeneous. Consequently, a product of $m$ bifundamentals and $s$ adjoint
insertions has degree $km+2s$ and $R$-charge
$\frac{(km+2s)}{(k+1)}$. Adding $\widetilde Q_i$ and $Q_j$ contributes $R(\widetilde Q_i)+R(Q_j)$ to the total $R$-charge, which corresponds to mesons of the electric theory.

\subsection{Independent meson operator construction}
\label{app:meson-basis-completeness}

In this section, we construct the basis \eqref{eq:meson-spectrum} of Section~\ref{sec:adeproposal}. The strategy of the construction is straightforward: we begin with the \(k=1\) results from~\cite{etingof2006centralextensionspreprojectivealgebras}, and then derive a basis for general \(k\) by substituting \(w=Z^k\).

 \textbf{Paths at $k=1$ modulo adjoint contribution}

To see the pure bifundamental paths, consider the quotient by setting the central adjoint element $Z$ to zero. 
The quotient 
\begin{equation}
    \Pi_0=\mathcal{B}_1/Z\mathcal{B}_1=\frac{\mathbb{C}\overline\Gamma}{ \langle\mu_1,\cdots,\mu_r\rangle}
\end{equation}
is called the ordinary preprojective algebra in \cite{etingof2006centralextensionspreprojectivealgebras}. The space $e_i\Pi_0 e_j$ consists of paths from $j$ to $i$ modulo relations $\mu_1=\cdots=\mu_r=0$. The Hilbert-series calculation \cite{etingof2006centralextensionspreprojectivealgebras} fixes the number of chains starting in degree $m$ at endpoints $(i,j)$ to be $$\dim(e_i\Pi_0e_j)_m=\nu_m^{ij},\qquad 0\leq m\leq h-2.$$ Since path classes span this space, choose $\nu_m^{ij}$ individual length-$m$ paths whose images form a basis:

 \begin{equation}\label{eq:selected-path}
B_{m,\alpha}^{ij}=X_{i_0i_1}\cdots X_{i_{m-1}i_m},\qquad i_0=i,\,i_m=j,\,1\leq\alpha\leq\nu_m^{ij}.     
 \end{equation}
At $m=0$, take $B_{0,1}^{ii}=e_i$.

\textbf{Adding the adjoint contribution }

The construction of~\cite{etingof2006centralextensionspreprojectivealgebras}
gives a complete homogeneous
Jordan basis of $\mathcal{B}_1$. At the endpoints $(i,j)$, it consists of $\nu_m^{ij}$ chains with an initial vector $v$ of degree $m$: 
\begin{equation}
v, Z v,\ldots,Z^{\ell_m-1}v,
\qquad
Z^{\ell_m} v=0,\qquad\ell_m=h-1-m.
\end{equation}
 Successive multiplication by $w$ adds an adjoint insertion.

Modulo $Z$, the selected paths \eqref{eq:selected-path} and the Jordan initial vectors $v$ are bases of the same space. In $\mathcal{B}_1$, a selected path can differ from $v$ by terms $Z^pv$, where $p>0$. Since $\deg Z=2$, $v$ starts a chain in degree $m-2p$. These lower-degree chains are $2p$ steps longer:
\begin{equation}
\ell_{m-2p}=h-1-(m-2p)=\ell_m+2p.
\end{equation}

Multiplying the path by $Z^q$ turns the correction into $Z^{p+q}v$. For $0\leq q<\ell_m$, it remains within that longer chain because
\[q+p\leq \ell_m-1+p<\ell_m+2p=\ell_{m-2p}.\]

Replacing the initial vectors $v$ with the selected paths of higher degree, therefore, adds only corrections from chains already retained. Thus, $Z^q B_{m,\alpha}^{ij}$ with $0\leq q\leq \ell_m-1$ forms a complete, linearly independent basis.

We can define the refined generating function for the basis: 

\begin{equation}
    F^{ij}(t,u)=\sum_{m=0}^{h-2}\nu_m^{ij}\sum_{q=0}^{h-2-m}t^{m+2q}u^q,
\end{equation}
where $t$ records the total
degree, while $u$ records the filtration index associated with powers of $w$. In particular,
\begin{equation}
F^{ij}(t,1) =\sum_{d\geq 0}\dim (e_i\mathcal B_1 e_j)_d t^d.
\end{equation}

This is the Hilbert series of $\mathcal B_1$.

\textbf{Extension to general $k$}

For general $k$, the defining F-term relations replace $w$ with
$Z^k$. Algebraically, this gives
\begin{equation}
\mathcal B_k
=
\mathcal B_1\otimes_{\mathbb C[w]}\mathbb C[Z],
\qquad
w=Z^k.
\end{equation}
Since $\mathbb C[Z]$ is free over $\mathbb C[w]$ with basis
$1,Z,\ldots,Z^{k-1}$, each $w$-chain becomes a $Z$-chain of length
\begin{equation}L_m=k\ell_m=k(h-1-m).\end{equation}

The extended basis consists of
\begin{equation}
   (w^q B_{m,\alpha}^{ij})\otimes Z^a\rightarrow Z^{kq+a}B_{m,\alpha}^{ij},\qquad 0\leq a\leq k-1,\,0\leq q\leq\ell_m-1 .
\end{equation}

 Thus $s=kq+a$ runs from $0$ to $k\ell_m-1$, giving
\begin{equation}\label{eq:vector-decomp}
V_k^{ij}
=
\bigoplus_{m=0}^{h-2}
\bigoplus_{\alpha=1}^{\nu_m^{ij}}
\bigoplus_{s=0}^{L_m-1}
\mathbb C\,Z^sB_{m,\alpha}^{ij}.
\end{equation}
Every path class has a unique expansion in these vectors.
Attaching $\widetilde Q_i$ and $Q_j$ gives the complete list of
meson species \eqref{eq:meson-spectrum}.

\subsection{Graded matching with the index prediction}
\label{app:meson-graded-matching}

The result above was derived from the electric F-terms and is independent of the large-$N$ index strategy. In this section, we will show that these two descriptions agree with one another, i.e.,
\begin{equation}\label{eq:hilb=P}\boxed{
  \operatorname{Hilb}(\mathcal B_k;y)=\mathcal{P}_{2kh}(y)} 
\end{equation}

The basis \eqref{eq:vector-decomp} gives the matrix Hilbert series\cite{etingof2006centralextensionspreprojectivealgebras}
\begin{equation}
\begin{aligned}
[\operatorname{Hilb}(\mathcal B_k;y)]_{ij}
&:=\sum_{d\geq0}\dim(V_k^{ij})_d\,y^d=\sum_{m=0}^{h-2}\nu_m^{ij}
  \sum_{s=0}^{k(h-1-m)-1}y^{km+2s},
\end{aligned}
\end{equation}
each coefficient counts operators of a definite $R$-charge in some ordered flavor channel.  

We use the identity appearing in the proof of
Corollary~4.6
of~\cite{etingof2006centralextensionspreprojectivealgebras},
 \begin{equation}\label{eq:given-Hilb}
\operatorname{Hilb}(\mathcal B_k;y)=\frac{1-y^{2kh}}{(1-y^2)(1-Gy^k+y^{2k})}   .
 \end{equation}
 
By the definition of $H(Q,t)$ \eqref{eq:graph-index-kernel-definition}, we obtain 
\begin{equation} \label{eq:hilb2}
 \frac{1-y^{2kh}}{(1-y^2)(1-Gy^k+y^{2k})} =  (1-y^{2kh})H(Q,y^{k+1}) . 
\end{equation}
By combining \eqref{eq:given-Hilb}, \eqref{eq:hilb2}, and the definition
$\mathcal P_L(y)=(1-y^L)H(Q,y^{k+1})$
from~\eqref{eq:ade-phys-kernel}, we prove the identity \eqref{eq:hilb=P}. Note that the large-$N$ index strategy supplies the family $\mathcal P_L(y)$, rather
than fixing $L$. The equality \eqref{eq:hilb=P} holds only when we choose $L = 2kh$.

\paragraph{Total multiplicities.}
In the Hilbert series \eqref{eq:given-Hilb}, setting $y=1$ counts the multiplicities of all species, and thus also counts the associated mesons,
\begin{equation}\frac{(1-y^{2kh})}{(1-y^2)}=1+y^2+\cdots+y^{2kh-2}\end{equation}
contributes $kh$, and the matrix $(1-G y^k+y^{2k})$ gives the Cartan matrix $C$. Hence
\begin{equation}
a_{ij}
=
\dim V_k^{ij}
=
kh(C^{-1})_{ij},
\qquad
C=2I_r-G.
\end{equation}
This recovers the proposed total multiplicities \eqref{eq:dual-rank} from the large-$N$ index strategy.

\subsection{Example:  \(A_2\) duality with adjoint }

For the $A_2$ Dynkin diagram, the rank and Coxeter number are $r=2$ and $h=3$. Only 
$m=0,1$ contributes to \eqref{eq:meson-spectrum}.
For \(m=0\), the starting paths are the identities, and the range of $s$ is
\[
0\le s\leq k(h-1)-1=2k-1.
\]
For \(m=1\), the starting paths are \(X_{12},X_{21}\), with
\[
0\le s\leq k(h-2)-1=k-1.
\]
Thus, the meson basis is summarized in Table~\ref{tab:a2-meson-spectrum}.

\begin{table}[H]
\centering
\renewcommand{\arraystretch}{1.3}
\begin{tabular}{c c c c}
\hline
Channel & Meson & Dressing range & Number of species \\
\hline
$(1,1)$
& $\widetilde Q_1X_1^sQ_1$
& $0\leq s\leq 2k-1$
& $2k$ \\

$(2,2)$
& $\widetilde Q_2X_2^sQ_2$
& $0\leq s\leq 2k-1$
& $2k$ \\

$(1,2)$
& $\widetilde Q_1X_1^sX_{12}Q_2$
& $0\leq s\leq k-1$
& $k$ \\

$(2,1)$
& $\widetilde Q_2X_2^sX_{21}Q_1$
& $0\leq s\leq k-1$
& $k$ \\
\hline
\end{tabular}
\caption{The $A_2$ meson spectrum obtained from
\eqref{eq:meson-spectrum}.}
\label{tab:a2-meson-spectrum}
\end{table}

Assigning degree two to each adjoint insertion and degree $k$
to each bifundamental, the matrix Hilbert series of the
internal meson paths is
\begin{equation}
\boxed{
\mathcal P_{6k}(y)=
\left(\sum_{s=0}^{k-1}y^{2s}\right)
\begin{pmatrix}
1+y^{2k}&y^k\\
y^k&1+y^{2k}
\end{pmatrix}.
}
\end{equation}
This agrees with the meson polynomial predicted by the
large-$N$ index strategy~\cite{Fang:2026ncq} and reproduces
the results in~\cite{Brodie:1996vx}.
\bibliographystyle{JHEP}
\bibliography{ref}
\end{document}